\documentclass[smallcondensed]{svjour3}  
\newif\ifworkinprogress
\pdfoutput=1

\usepackage[utf8]{inputenc}

\usepackage{microtype}
\usepackage{bm}
\usepackage{amssymb,amsmath}
\usepackage{graphicx}
\usepackage[center]{subfigure}
\usepackage{algpseudocode}\usepackage{xspace}

\usepackage{color}

\makeatletter
\def\cl@chapter{\@elt {theorem}}
\makeatother
\usepackage{cleveref} 

\usepackage{url}
\usepackage{units}
 
\usepackage{mathptmx}

\usepackage[ruled]{algorithm2e}

\usepackage{tikz}
\usetikzlibrary[shapes,arrows,arrows.meta]
\usetikzlibrary{positioning}
\usetikzlibrary{fit}
\usetikzlibrary{matrix}
\usetikzlibrary{calc,positioning,decorations.pathreplacing}
\usetikzlibrary{bayesnet}
\pgfdeclarelayer{bg}\pgfsetlayers{bg,main}

\newcommand*\circled[1]{\tikz[baseline=(char.base)]{
		\node[shape=circle,draw,inner sep=1pt] (char) {#1};}}

\newcommand{\approachname}{\mbox{MixedTrails}\xspace}

\newif\ifworkinprogressomments
\ifworkinprogress
    \newcommand{\phsi}[1]{\textbf{\color{red}/* #1 (phsi) */}}
    \newcommand{\mbe}[1]{\textbf{\color{blue}/* #1 (mbe) */}}
    \newcommand{\fle}[1]{\textbf{\color{brown}/* #1 (fle) */}}
    \newcommand{\mst}[1]{\textbf{\color{olive}/* #1 (mst) */}}
    \newcommand{\aho}[1]{\textbf{\color{olive}/* #1 (aho) */}}
\else
    \newcommand{\phsi}[1]{}
    \newcommand{\mbe}[1]{}
    \newcommand{\fle}[1]{}
    \newcommand{\mst}[1]{}
    \newcommand{\aho}[1]{}
\fi

\newenvironment{revised}{}{\ignorespacesafterend}

\newcommand{\para}[1]{\smallskip\noindent\textbf{#1}}

\newcommand\numberthis{\addtocounter{equation}{1}\tag{\theequation}}

\newcommand{\furl}[1]{\footnote{\url{#1}}}

\newcommand{\mtheta}[0]{\bm{\theta}}
\newcommand{\malpha}[0]{\bm{\alpha}}
\newcommand{\mgamma}[0]{\bm{\gamma}}
\newcommand{\mphi}[0]{\bm{\phi}}

\newcommand{\vtheta}[0]{\bm{\theta}}
\newcommand{\valpha}[0]{\bm{\alpha}}
\newcommand{\vgamma}[0]{\bm{\gamma}}

\newcommand{\vphi}[0]{\bm{\phi}}
\newcommand{\vn}[0]{\bm{n}}
\newcommand{\mn}[0]{\bm{n}}

\newcommand{\vkappa}[0]{\bm{\kappa}}

\title{MixedTrails: Bayesian Hypothesis Comparison \\ on Heterogeneous Sequential Data}

\author{Martin Becker \and Florian Lemmerich \and Philipp Singer \and Markus Strohmaier \and Andreas Hotho}
\authorrunning{M. Becker, F. Lemmerich, P. Singer, M. Strohmaier, A. Hotho}

\institute{Martin Becker and Andreas Hotho \at
              Data Mining and Information Retrieval Group, University of Würzburg, Würzburg, Germany \\
              \email{\{becker, hotho\}@informatik.uni-wuerzburg.de} 
           \and
           Florian Lemmerich and Philipp Singer and Markus Strohmaier \at
        GESIS - Leibniz Institute for the Social Sciences,
Cologne, Germany\\
\email{\{florian.lemmerich, philipp.singer, markus.strohmaier\}@gesis.org}
}

\date{July 2017}

\begin{document}

\maketitle

\begin{abstract}
Sequential traces of user data are frequently observed online and offline, e.g., as sequences of visited websites or as sequences of locations captured by GPS.
However, understanding factors explaining the production of sequence data is 
a challenging task, especially since the data generation is often not homogeneous. For example, navigation behavior might change in different phases of browsing a website, or movement behavior may vary between groups of users.
In this work, we tackle this task and propose \emph{\approachname{}}, a Bayesian approach for comparing the plausibility of hypotheses regarding the generative processes of heterogeneous sequence data.
Each hypothesis is derived from existing literature, theory or intuition and represents a belief about transition probabilities between a set of states that can vary between groups of observed transitions. 
For example, when trying to understand human movement in a city and given some observed data, a hypothesis assuming tourists to be more likely to move towards points of interests than locals, can be shown to be more plausible than a hypothesis assuming the opposite.
Our approach incorporates such hypotheses as Bayesian priors in a generative mixed transition Markov chain model, and compares their plausibility utilizing Bayes factors.
We discuss analytical and approximate inference methods for calculating the marginal likelihoods for Bayes factors, give guidance on interpreting the results, and illustrate our approach with several experiments on synthetic and empirical data from Wikipedia and Flickr.
Thus, this work enables a novel kind of analysis for studying sequential data in many application areas.

\end{abstract}
\begin{figure}[t!]
	\centering
	\begin{tikzpicture}[>={Triangle[width=2mm,length=2mm]}]
\node(graph)[] {
	\begin{tikzpicture}[]
    	\node[] (n1) {\includegraphics[scale=0.03]{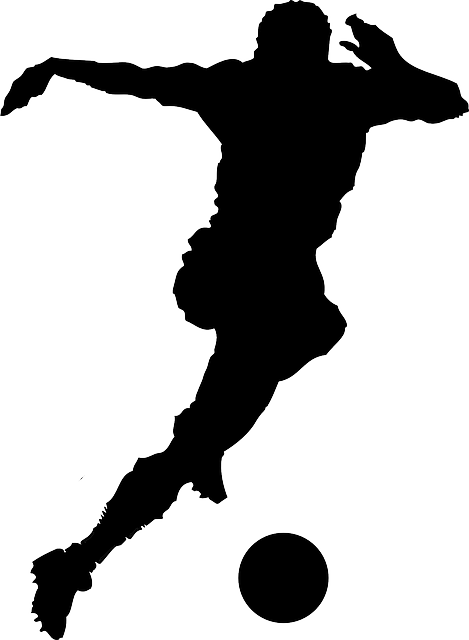} \circled{1}};
    	\node[right = of n1] (n2) {\includegraphics[scale=0.03]{res_soccer} \circled{2}};
    	\node[above = of n1] (n3) {\includegraphics[scale=0.03]{res_soccer} \circled{3}};
    	\node[right = of n3] (n4) {\includegraphics[scale=0.03]{res_soccer} \circled{4}};
    	\coordinate (Middle) at ($(n3)!0.5!(n4)$);
    	\node[above = of Middle] (n5) {\includegraphics[scale=0.02]{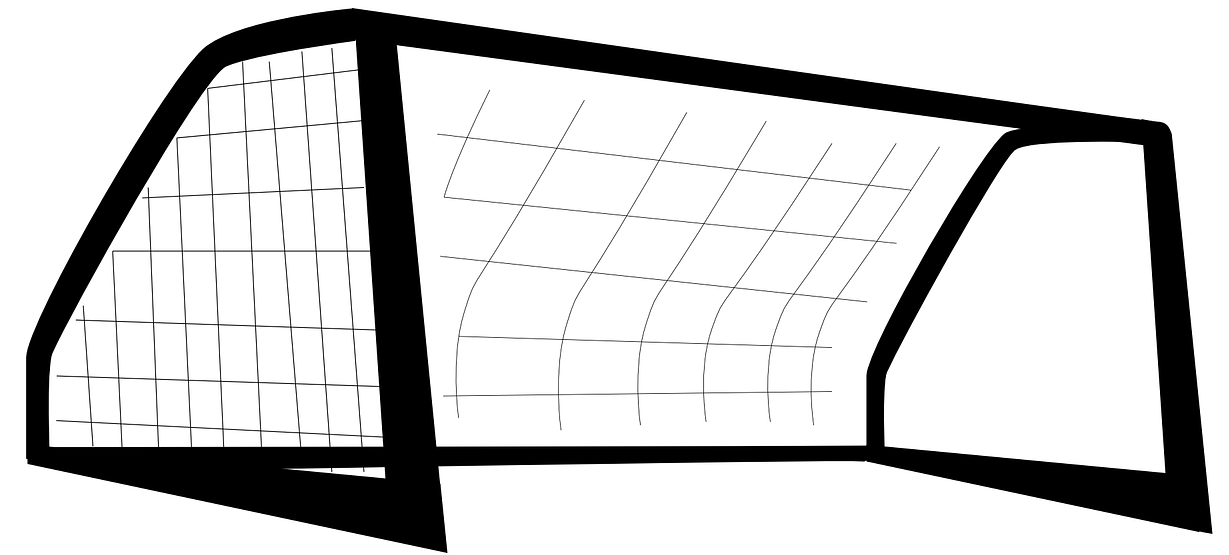} \circled{5}};
    	
    	\draw[->, color=black] (n1.-25) -- (n2.205) node [midway, below] {20};
    	\draw[->, color=black] (n1) -- (n3) node [midway, left] {20};
    	\draw[->, color=black] (n2) -- (n1)  node [midway, above] {20};
    	\draw[->, color=black] (n2) -- (n4) node [midway, right] {20};
    	\draw[->, color=black] (n3.-65) -- (n1.65) node [midway, right] {10};
    	\draw[->, color=black] (n3.-25) -- (n4.205) node [midway, below] {10};
    	\draw[->, color=black] (n3) -- (n5) node [midway, left] {20};
     	\draw[->, color=black] (n4.-115) -- (n2.115)node [midway, left] {10};
    	\draw[->, color=black] (n4) -- (n3) node [midway, above] {10};
    	\draw[->, color=black] (n4) -- (n5) node [midway, right] {20};
	\end{tikzpicture}
};
\matrix (table)[matrix of nodes, left = 0.3cm of graph.south west, yshift=0.1cm, anchor=south east, inner sep=2.5pt,outer sep=0, minimum height=0.2cm,  minimum width=0.5cm,
column 1/.style={nodes={text width=1.3cm}}, column 2/.style={nodes={text width=1.3cm}}] {
    Kicker & Receiver & Min. \\
    Player (1) & Player (3) & \node[color=red]{5}; \\
    Player (3) & Goal   (5) & \node[color=red]{7}; \\
    Player (3) & Goal   (5) & \node[color=red]{8}; \\
    Player (2) & Player (4) & \node[color=red]{17}; \\
    ... & ... & ... \\
    Player (4) & Player (3) & \node[color=blue]{85}; \\
    Player (2) & Player (1) & \node[color=blue]{89}; \\
};
\draw[] (table-1-1.south west) -- (table-1-3.south east);
\node(graph_groups)[right = 0.3cm of graph.south east, anchor=south west] {
	\begin{tikzpicture}[]
    	\node[] (n1) {\includegraphics[scale=0.03]{res_soccer} \circled{1}};
    	\node[right = of n1] (n2) {\includegraphics[scale=0.03]{res_soccer} \circled{2}};
    	\node[above = of n1] (n3) {\includegraphics[scale=0.03]{res_soccer} \circled{3}};
    	\node[right = of n3] (n4) {\includegraphics[scale=0.03]{res_soccer} \circled{4}};
    	\coordinate (Middle) at ($(n3)!0.5!(n4)$);
    	\node[above = of Middle] (n5) {\includegraphics[scale=0.02]{res_goal} \circled{5}};
    	\node[left = .2 of n5, align=center, color=red] (1st) {first\\half};
    	\node[right = -0.1 of n5, align=center, color=blue] (2st) {second\\half};
    	
    	\draw[->, color=blue] (n1.-25) -- (n2.205) node [midway, below] {20};
    	\draw[->, color=red, dashed] (n1) -- (n3) node [midway, left] {20};
    	\draw[->, color=blue] (n2) -- (n1)  node [midway, above] {20};
    	\draw[->, color=red, dashed] (n2) -- (n4) node [midway, right] {20};
    	\draw[->, color=blue] (n3.-65) -- (n1.65) node [midway, right] {10};
    	\draw[->, color=blue] (n3.-25) -- (n4.205) node [midway, below] {10};
    	\draw[->, color=red, dashed] (n3) -- (n5) node [midway, left] {20};
     	\draw[->, color=blue] (n4.-115) -- (n2.115)node [midway, left] {10};
    	\draw[->, color=blue] (n4) -- (n3) node [midway, above] {10};
    	\draw[->, color=red, dashed] (n4) -- (n5) node [midway, right] {20};
    	
    	\draw[->, color=red, dashed, , shorten <=0.1cm, shorten >=0.0cm]
    	    (1st.south west) -- (1st.south east); 
    	\draw[->, color=blue, shorten <=0.25cm, shorten >=0.2cm] 
    	    (2st.south west) -- (2st.south east);
	\end{tikzpicture}
};

\draw [decoration={brace},decorate,line width=1pt] ([yshift=-0.3cm]table.south east) -- ([yshift=-0.3cm]table.south west) 
node [midway, below, yshift=-0.1cm] {(a) transitions};
\draw [decoration={brace},decorate,line width=1pt] ([yshift=-0.2cm]graph.south east) -- ([yshift=-0.2cm]graph.south west) 
node [midway, below, yshift=-0.1cm] {(b) transition graph};

\draw [decoration={brace},decorate,line width=1pt] ([yshift=-0.2cm]graph_groups.south east) -- ([yshift=-0.2cm]graph_groups.south west) 
node [midway, below, yshift=-0.1cm] {(c) heterogeneous transitions};

\node[fit=(graph)(table)(graph_groups)] (row1) {};

\node(hyp_off)[below left = 1.5cm and 0.25cm of row1.south] {
	\begin{tikzpicture}[]
    	\node[] (n1) {\includegraphics[scale=0.03]{res_soccer}};
    	\node[right = of n1] (n2) {\includegraphics[scale=0.03]{res_soccer}};
    	\node[above = of n1] (n3) {\includegraphics[scale=0.03]{res_soccer}};
    	\node[right = of n3] (n4) {\includegraphics[scale=0.03]{res_soccer}};
    	\coordinate (Middle) at ($(n3)!0.5!(n4)$);
    	\node[above = of Middle] (n5) {\includegraphics[scale=0.02]{res_goal}};
    	
    	\draw[->, color=gray, line width=2.5pt] (n1) -- (n3);
    	\draw[->, color=gray] (n1) -- (n4);
    	\draw[->, color=gray] (n2) -- (n3);
    	\draw[->, color=gray, line width=2.5pt] (n2) -- (n4);
    	\draw[->, color=gray, line width=2.5pt] (n3) -- (n5);
    	\draw[->, color=gray, line width=2.5pt] (n4) -- (n5);

	\end{tikzpicture}
};

\node(hyp_left)[below right = 1.5cm and 0.25cm of row1.south] {
	\begin{tikzpicture}[]
	\node[] (n1) {\includegraphics[scale=0.03]{res_soccer}};
	\node[right = of n1] (n2) {\includegraphics[scale=0.03]{res_soccer}};
	\node[above = of n1] (n3) {\includegraphics[scale=0.03]{res_soccer}};
	\node[right = of n3] (n4) {\includegraphics[scale=0.03]{res_soccer}};
	\coordinate (Middle) at ($(n3)!0.5!(n4)$);
	\node[above = of Middle] (n5) {\includegraphics[scale=0.02]{res_goal}};
	
	\draw[->, color=gray, line width=2.5pt] (n2) -- (n1);
	\draw[->, color=gray, line width=2.5pt] (n1) -- (n3);
	\draw[->, color=gray, line width=2.5pt] (n4) -- (n3);
	\draw[->, color=gray, line width=2.5pt] (n3) -- (n5);

	\end{tikzpicture}
};

\node(hyp_uni)[left = 0.5cm of hyp_off.north west, anchor=north east, xshift=0cm] {
	\begin{tikzpicture}[]
    	\node[] (n1) {\includegraphics[scale=0.03]{res_soccer}};
    	\node[right = of n1] (n2) {\includegraphics[scale=0.03]{res_soccer}};
    	\node[above = of n1] (n3) {\includegraphics[scale=0.03]{res_soccer}};
    	\node[right = of n3] (n4) {\includegraphics[scale=0.03]{res_soccer}};
    	\coordinate (Middle) at ($(n3)!0.5!(n4)$);
    	\node[above = of Middle] (n5) {\includegraphics[scale=0.02]{res_goal}};
    	\draw[->, color=gray] (n1) -- (n2);
    	\draw[->, color=gray] (n1) -- (n3);
    	\draw[->, color=gray] (n1) -- (n4);
    	\draw[->, color=gray] (n1) -- (n5);
    	\draw[->, color=gray] (n2) -- (n1);
    	\draw[->, color=gray] (n2) -- (n3);
    	\draw[->, color=gray] (n2) -- (n4);
    	\draw[->, color=gray] (n2) -- (n5);
    	\draw[->, color=gray] (n3) -- (n1);
    	\draw[->, color=gray] (n3) -- (n2);
    	\draw[->, color=gray] (n3) -- (n4);
    	\draw[->, color=gray] (n3) -- (n5);
    	\draw[->, color=gray] (n4) -- (n1);
    	\draw[->, color=gray] (n4) -- (n2);
    	\draw[->, color=gray] (n4) -- (n3);
    	\draw[->, color=gray] (n4) -- (n5);
	\end{tikzpicture}
};

\node(hyp_def)[right = 0.5cm of hyp_left] {
	\begin{tikzpicture}[]
    	\node[] (n1) {\includegraphics[scale=0.03]{res_soccer}};
    	\node[right = of n1] (n2) {\includegraphics[scale=0.03]{res_soccer}};
    	\node[above = of n1] (n3) {\includegraphics[scale=0.03]{res_soccer}};
    	\node[right = of n3] (n4) {\includegraphics[scale=0.03]{res_soccer}};
    	\coordinate (Middle) at ($(n3)!0.5!(n4)$);
    	\node[above = of Middle] (n5) {\includegraphics[scale=0.02]{res_goal}};
    	\draw[->, color=gray, line width=2.5pt] ($(n1)+(.5,0)$) -- (n2);
    	\draw[->, color=gray, line width=2.5pt] ($(n2)-(.5,0)$) -- (n1);
    	\draw[->, color=gray] (n3) -- (n1);
    	\draw[->, color=gray, line width=2.5pt] ($(n3)+(.5,0)$) -- (n4);
    	\draw[->, color=gray] (n4) -- (n2);
    	\draw[->, color=gray, line width=2.5pt] ($(n4)-(.5,0)$) -- (n3);
    	
	\end{tikzpicture}
};

\draw [decoration={brace},decorate,line width=1pt] ([yshift=0.0cm]hyp_uni.north west) -- ([yshift=0.0cm]hyp_uni.north east) 
node [midway, above, yshift=0.1cm] {(d) uniform hypothesis};

\draw [decoration={brace},decorate,line width=1pt] ([yshift=0.0cm]hyp_off.north west) -- ([yshift=0.0cm]hyp_off.north east) 
node [midway, above, yshift=0.1cm] {(e) offense hypothesis};

\draw [decoration={brace},decorate,line width=1pt] ([yshift=0.0cm]hyp_left.north west) -- ([yshift=0.0cm]hyp_left.north east) 
node [midway, above, yshift=0.1cm] {(f) left-flank hypothesis};

\draw [decoration={brace},decorate,line width=1pt] ([yshift=0.0cm]hyp_def.north west) -- ([yshift=0.0cm]hyp_def.north east) 
node [midway, above, yshift=0.1cm] {(g) defense hypothesis};

\end{tikzpicture}

	\caption{\textbf{Illustrating example. } 
		In this figure, we show an illustrating soccer example: We are interested in a team's strategy in a specific game. We start with data on passes and shots (a). Using a simple Markov chain, we can model these as transitions between states (b). The previously proposed HypTrails approach allows researchers to compare homogeneous hypotheses about sequential data that express beliefs in transition probabilities (d-g, strength of belief indicated by line width). Utilizing Bayesian inference, it then determines the evidence of the data (b) under these hypotheses (d-g) and ranks the hypotheses based on their plausibility; in this case, the uniform hypothesis (d) is the relatively most plausible one. 
		However, HypTrails is limited to homogeneous data, and does not allow for more fine-grained hypotheses. Indeed, (c) reveals that splitting the data into halftimes allows for a significantly better explanation of the data: A hypothesis that assumes offense (e) in the first halftime and defense (g) in the second appears to be a lot more plausible. \approachname{} enables the comparison of such hypotheses on heterogeneous data.
}	
	\label{fig:figure1}
\vspace{-2em}
\end{figure}

\vspace{-5mm}
\section{Introduction}
Sequential data over a discrete state space emerges in a variety of settings, including sequences of weather conditions \cite{gabriel1962markov}, DNA sequences \cite{smith1985fluorescence}, Web navigation \cite{page1999pagerank}, or real-world travel sequences over locations \cite{gonzalez2008understanding,noulas2012tale}.
Understanding the underlying processes that generate such sequences can be useful for a wide range of applications, such as improving network structures, predicting user clicks on websites, or enhancing recommendations, and has been a challenge and complex research objective in our community for years. 

\para{Background. }
The (first-order) Markov chain model is one of the most elementary, yet versatile, models for transitions between sequence states. 
It follows the Markovian assumption that the probability of the next state in a sequence depends exclusively on the current state. 
Building upon this basic model, 
the recently proposed HypTrails approach~\cite{singer2015} allows to compare hypotheses about sequential data, where
hypotheses represent beliefs in state transition probabilities that are derived from existing literature, theory, or intuition with regard to the respective application domain. 
For example, by studying Wikipedia user data, we found that the hypothesis that users preferably click on links at the top of a page provides a better explanation of user navigation than a hypothesis that assumes transitions to semantically similar pages~\cite{dimitrov2016}.

\Cref{fig:figure1} shows a concrete example on soccer data. It features passes between players and shots at the goal (a).
In this scenario, we are interested in the strategy a team has used in a game, e.g., an offensive strategy, a defensive strategy, or just random passing.
For this purpose, we construct a Markov transition model using the players and the goal as states, and the passes and shots as transitions between these states (b).
With HypTrails~\cite{singer2015}, researchers can then express and compare hypotheses (d-g) about pass sequences by specifying different beliefs in transitions. 
For instance, a simple hypothesis could state that all transitions are equally likely (d). 
Other hypotheses may express predominance of 
offensive passing (e), 
a left-flank strategy (f),
or defensive play (g).
Given such hypotheses, HypTrails calculates the Bayesian evidence of the data under each hypothesis based on which we can rank their relative plausibility. Given the transition data in \Cref{fig:figure1}(a), the approach would rank the uniform hypothesis (d) as the most plausible one, as it resembles the overall data (b) best.

\para{Problem and objectives. }
Simple Markov chain models, and consequently also the HypTrails approach, assume homogeneous sequence data. As such, they cannot take into account behavior stemming from several underlying processes. 
For instance, research on mobility has found starkly differing user groups
such as tourists and locals~\cite{lemmerich2016subtrails}, and there exist different phases of Web navigation with distinct patterns \cite{west2012human}. Reconsidering our soccer scenario of \Cref{fig:figure1}, we can observe that the play style substantially differs for the 1st and 2nd half of the game (dashed and solid arrows).
As a consequence, a hypothesis that assumes offensive play for the first halftime and defensive play for the second halftime (cf. \Cref{fig:figure1} (e) and (g)) could provide a better explanation for our data, but cannot be compared with existing approaches.

To that end, our goal in this paper is to propose a method that lets researchers intuitively formalize and compare hypotheses about heterogeneous sequence data, such as ``The team played according to the offense hypothesis in the first halftime, and according to the defense hypothesis in the second halftime.'' 
In this context, we aim at a general and flexible approach: 
allowing to group transitions by a variety of features, like user groups,  state properties, or the set of antecedent transitions on the one hand, and enabling users to formulate probabilistic group assignments as in the context of smooth behavioral shifts or uncertain classifiers on the other hand.

\para{Contributions. }
In this paper, we introduce the \emph{\approachname{}} approach, which covers all necessary aspects to enable the comparison of hypotheses on heterogeneous sequence data:
(i) We suggest a method to formalize hypotheses as belief matrices and probabilistic group memberships; 
(ii) We propose the Mixed Transition Markov Chain (MTMC) model that allows to capture such hypotheses; 
(iii) We show how to elicit priors for this model according to the given hypotheses; 
(iv) We discuss exact and approximate inference for our model; 
(v) We provide guidance in the interpretation of the result plots.
Finally, we demonstrate the benefits of our approach with synthetic and real world datasets.

Overall, we present a novel approach for specifying and comparing hypotheses about heterogeneous sequence data that involve varying behavior in parts of the observed transitions.
This will enable researchers and practitioners to perform a new kind of analysis on such data.

\vspace{-4mm}
\section{Background and Notation}
\label{sec:background}
\vspace{-2mm}
In this section, we shortly introduce HypTrails~\cite{singer2015} which our approach \emph{\approachname{}} builds on and cover its building blocks, i.e., Markov chains and Bayesian model comparison. An overview of all important notations used throughout this article can be found in \Cref{app:notation}.

\para{Markov Chain Model. }
\label{sec:mcm}
A Markov chain model $M_{MC}$ \cite{kemeny1960finite,singer2014detecting} is a random process modeling a sequence of random variables $X_1,X_2,\ldots,X_l$ as transitions between a set of states $S=\{s_1,s_2,\ldots,s_n\}$.
In this work, we focus on first-order Markov chains, which describe a memoryless process meaning that the next state $s_j$ in a sequence only depends on the current one $s_{i_\tau}$, i.e.:
$\Pr(X_{\tau+1}\,=\,s_j|\,X_1\,=\,s_{i_1},\ldots,X_\tau\,=\,s_{i_\tau}) 
=\Pr(X_{\tau+1}\,=\,s_j\,|\,X_\tau\,=\,s_i) = \theta_{i,j}$.
The parameters of a Markov chain model $M_{MC}$ are the transition probabilities $\theta_{i,j}$ between states $s_i$ and $s_j$ represented by a transition matrix $\mtheta = (\theta_{i,j})$.
As the model is stochastic, each row of the transition matrix sums to $1$, i.e. $\forall i: \sum_j \theta_{i,j} = 1$. 
Thus, given a transition dataset $D$ with transition counts $n_{i,j}$ between states $s_i$ and $s_j$, the likelihood of observing these transitions is:
\begin{equation*}
\Pr(D|\mtheta,M_{MC}) = \prod_{t_k \in D} \theta_{i_k,j_k} = \prod_{s_i,s_j \in S} \theta_{i,j}^{n_{i,j}}
\label{eq:likelihoodmcm}
\end{equation*}
\begin{revised}
In the Bayesian setting, \emph{prior} beliefs in transition probabilities are updated after observing data. The choice of the prior distribution over the transition probabilities $\mtheta$ is crucial for calculating the posterior or examining the marginal likelihood.
As detailed below, in this paper, we employ independent Dirichlet priors for each state $i$, i.e., $\vtheta_{s_i} \sim Dir(\valpha_{s_i})$ where $\valpha_{s_i}$ are the parameters of the Dirichlet distribution.

\end{revised}

\para{Bayesian model comparison. }
\label{sec:modelcomparison}
Given a set of models $\{M_1, \ldots, M_m\}$ and some data $D$, Bayesian model comparison establishes a partial order on the set of models $M_i \sqsubseteq M_j \sqsubseteq M_k$ based on the marginal likelihood $Pr(D|M_i)$ of the data $D$ given each model $M_i$. 
The marginal likelihood represents the plausibility of the model.
\begin{revised}
The strength of evidence in favor of a model $M_i$ compared to a model $M_j$ can then be formally measured by a Bayes factor $B_{i,j}$~\cite{kass1995bayes}.
It represents the factor by which the prior odds in favor of one of two compared models change after seeing the data (posterior odds):
\vspace{-1mm}
\begin{align}
\underbrace{\frac{\Pr(M_i|D)}{\Pr(M_j|D)}}_{\text{posterior odds}} = B_{i,j} \cdot \underbrace{\frac{\Pr(M_i)}{\Pr(M_j)}}_{\text{prior odds}}, ~\text{with } \underbrace{B_{i,j} = \frac{\Pr(D|M_i)}{\Pr(D|M_j)}}_{\text{Bayes factor}}
\label{eq:bayesfactor}
\end{align}
\end{revised}

Bayes factors can also be utilized for conducting Bayesian hypotheses comparison if priors encode theory-induced hypotheses as advocated in \cite{kruschke2014doing,rouder2009bayesian,vanpaemel2010prior}; we use this throughout this article.
For judging significance, we refer to Kass and Raftery's interpretation table~\cite{kass1995bayes}.

\para{HypTrails. }
\label{sec:hyptrails}
HypTrails~\cite{singer2015} operationalizes Bayesian model comparison for hypotheses on Markov chain models $M_{MC}$ in order to establish a partial order $\sqsubseteq$ on a set of hypotheses $\mathcal{H} = \{H_1, \ldots, H_n\}$ based on their plausibility given the data.
A hypothesis $H$ is expressed as a prior probability distribution $\Pr(\mtheta|H,M_{MC})$ over all instances of transition probability matrices $\mtheta$, which is required to compute the marginal likelihood used by the Bayes factor:
\begin{align*}
\underbrace{\Pr(D|H,M_{MC})}_{\text{marginal likelihood}} = \int \underbrace{\Pr(D|\mtheta,M_{MC})}_{\text{likelihood}} ~ \underbrace{\Pr(\mtheta|H,M_{MC})}_{\text{prior}} ~ d\mtheta
\end{align*}
\begin{revised}
If we now assume all hypotheses to be equally likely  a-priori (as often done in Bayesian model comparison), the Bayes factor directly implies the posterior probabilities, cf. the derivation of Bayes factor in \cite{kass1995bayes}.
\end{revised}

To express a hypothesis $H$ about transition probabilities $\Pr(\mtheta|H,M_{MC})$, HypTrails uses Dirichlet priors, i.e., for each state $s_i$ an individual Dirichlet prior $Dir(\valpha_{s_i})$ is specified which defines beliefs about transition probabilities from that state $s_i$ to all other states.
The parameters $\valpha_{s_i}$ are vectors of positive numbers, i.e.,
$\valpha_{s_i} = (\alpha_{i,1}, \ldots \alpha_{i,n}), ~ \alpha_{i,j} \in \mathbb{R}^+$.
That is, given a hypothesis $H$ and a fixed number of imaginary (pseudo) transitions originating from state $s_i$, $\alpha_{i,j} - 1$ denotes the expected number of observed transitions from state $s_i$ to state $s_j$.

The process of expressing a belief as a formal hypothesis $H$ and transforming it into prior parameters (pseudocounts) is called \emph{elicitation}.
For elicitation, HypTrails assumes a two step process:
First, a transition probability distribution $\vphi_{s_i} = (\phi_{i,1}, \ldots, \phi_{i,n})$ is specified for each state $s_i$, resulting in a stochastic transition matrix $\vphi = (\phi_{i,j})$.
Then, a concentration factor $\kappa \in \mathbb{N}^+_0$ is set in order to derive the hyperparameters $\alpha_{i,j}$ by calculating: $\malpha = \kappa\ \cdot \mphi + 1$,
where $\kappa$ is proportional to the amount of pseudocounts we assign to each state
\footnote{\revised Note that this is a slightly simplified version of the original Trial Roulette method from the HypTrails paper \cite{singer2015} regarding two aspects. First, we do not distribute chips but multiply by a concentration factor which is effectively equivalent and easier to compute. Second, we assume in this paper the same weight in each row of the Markov chain which makes formulating hypotheses and interpreting results easier. However, these simplifications are not required and reverting them is straightforward.}. 
The $+1$ adds the proto-prior that is necessary to ensure proper priors. 
Also, if $\kappa = 0$, every transition probability configuration is equally likely (referred to as a flat prior, cf.~\cite{singer2015}).
The higher we set the concentration factor $\kappa$, the more we ``believe'' in our hypothesis, i.e., we get higher marginal likelihood values if we are correct, but we are also penalized more if the hypothesis is off.
The lower we set the concentration factor $\kappa$, the more ``slack'' we allow for our hypothesis, i.e., we are not as strongly penalized for errors, but we also cannot reach large marginal likelihood values if we are correct. 
\begin{revised}
Note that in the general framework of Bayesian model comparison, choosing priors for the corresponding model parameters is not an easy task, since usually a variety of information has to be taken into account including relevant data, literature, or certainty in the belief. 
HypTrails somewhat alleviates this issue by 
formalizing the suggestion from \cite{kass1995bayes} to compare several prior instantiations by using a range of concentration factors $\vkappa = \{\kappa_1, \kappa_2, ...\}$.
This allows for a structured and detailed comparison of hypotheses.
Also see \Cref{sec:example} for a discussion on alternative approaches.
\end{revised}

\section{\approachname{}: Bayesian Hypothises Comparison in Heterogeneous Sequence Data}
\label{sec:methodology}
In this section, we introduce our approach \approachname{} for comparing hypotheses about heterogeneous sequence data using Bayesian model comparison.
To this end, we first elaborate on the specific problem setting (Section~\ref{sec:problem}) and explain how hypotheses for heterogeneous sequence data are structured. 
Then, we introduce the Mixed Transition Markov Chain (MTMC) model (Section~\ref{sec:model}) --- an extension of the basic Markov chain model --- that allows to model such heterogeneous data. 
By incorporating hypotheses as elicited priors over the model parameters (Section~\ref{sec:eliciting}), we can utilize Bayesian model comparison to make relative judgements about the plausibility of the given hypotheses.
Finally, we derive an approach for model inference (Section~\ref{sec:inference}) and give guidelines for interpreting the results (Section~\ref{sec:example}). 
For illustrative purposes, we will refer to the soccer example visualized in \Cref{fig:figure1}.

\subsection{Problem statement and approach}
\label{sec:problem}
The goal of this paper is to compare hypotheses about \emph{heterogeneous} sequence data. 
That is, considering a dataset of transitions $D = \{t_1, \ldots, t_m\}$ between a set of states $S=\{s_1, \ldots s_n\}$, we want to establish a partial ordering $\sqsubseteq$ on a set of given hypotheses $\mathcal{H} = \{H_1, H_2, \ldots\}$ that express \emph{how} the observed transitions may have been generated.
Extending HypTrails~\cite{singer2015}, we focus on 
transitions generated by several independent processes.

\para{Hypotheses}
We describe a heterogeneous hypothesis $H=(\mgamma, \mphi)$ by two components.
First, the \textit{group assignment probabilities} $\mgamma$ associate each transition $t \in D$ in the dataset $D$ with a probability distribution $\mgamma_t$ over a set of \textit{groups} $G = \{g_1, \ldots, g_o\}$ defined by the corresponding hypothesis.
We write all group assignment probabilities for a hypothesis as $\mgamma = \{ \vgamma_t | t \in D\}$, with $\mgamma_t = \{\gamma_{g|t}|g \in G\}$. Here, $\gamma_{g|t}$ is the probability that transition $t$ belongs to group $g$.
Second, the \textit{group transition probabilities} $\mphi$ describe the behavior of each group $g \in G$ by specifying respective transition probabilities between states.
Formally, all group transition probabilities according to a given hypotheses are written as $\mphi = (\mphi_1, ...\mphi_o)$, with $\mphi_g = (\phi_{i,j|g})$, where $\phi_{i,j|g}$ is the probability of observing a transition to state $s_j$ given state $s_i$ within group $g$.
Note that a \emph{homogeneous} hypothesis can be regarded as a special case of a 
heterogeneous one where all transition are assigned deterministically
to one group.

\para{Comparison}
Given several hypotheses, \approachname{} --- just like HypTrails --- establishes a partial order $\sqsubseteq$ by employing Bayes factors to compare their relative plausibility with respect to a dataset $D$.
This is done by converting each hypothesis $H_i$ into Bayesian priors (see \Cref{sec:eliciting}) of the generative model MTMC (see \Cref{sec:model}) and calculating the marginal likelihood (Bayesian evidence).

\begin{figure}
    \begin{tikzpicture}[]

\matrix (p_gt_0)[matrix of nodes, inner sep=2.5pt,outer sep=0, minimum height=0.2cm,  minimum width=0.5cm,
column 1/.style={nodes={text width=1.3cm}}, 
column 2/.style={nodes={text width=1.7cm}}] {
    Kicker & Receiver & $\gamma_{1|t}$ \\
    Player (1) & Player (3); & \textbf{1.0} \\
    Player (3) & Player (5) & \textbf{1.0} \\
    Player (3) & Goal   (5) & \textbf{1.0} \\
    Player (2) & Player (4) & \textbf{1.0} \\
    ... & ... & ... \\
    Player (4) & Goal   (2) & \textbf{1.0} \\
    Player (2) & Player (1) & \textbf{1.0} \\
};
\draw[] (p_gt_0-2-1.north west) -- (p_gt_0-2-3.north east);

\matrix (theta_uniform)[matrix of math nodes,left delimiter={(},right delimiter={)}, above right = -0.5cm and 3.2cm of p_gt_0, anchor=north west, inner sep=.5pt,outer sep=0, minimum height=0.4cm,  minimum width=0.4cm
] {
	0 & \textbf{\nicefrac{1}{4}} & \textbf{\nicefrac{1}{4}} & \textbf{\nicefrac{1}{4}} & \textbf{\nicefrac{1}{4}} \\
	\textbf{\nicefrac{1}{4}} & 0 & \textbf{\nicefrac{1}{4}} & \textbf{\nicefrac{1}{4}} & \textbf{\nicefrac{1}{4}} \\
	\textbf{\nicefrac{1}{4}} & \textbf{\nicefrac{1}{4}} & 0 & \textbf{\nicefrac{1}{4}} &\textbf{ \nicefrac{1}{4}} \\
	\textbf{\nicefrac{1}{4}} & \textbf{\nicefrac{1}{4}} & \textbf{\nicefrac{1}{4}} & 0 & \textbf{\nicefrac{1}{4}} \\
};
\draw [decoration={brace},decorate,line width=1pt] ([yshift=-0.2cm]theta_uniform.south east) -- ([yshift=-0.2cm]theta_uniform.south west) 
node [midway, below, yshift=-0.1cm] {$\mphi_\text{uniform}$};

\matrix (pgthalves)[matrix of nodes, below = 0.5cm of p_gt_0.south west, xshift=0cm, anchor=north west, inner sep=2.5pt,outer sep=0, minimum height=0.2cm,  minimum width=0.5cm,
column 1/.style={nodes={text width=1.3cm}}, 
column 2/.style={nodes={text width=1.3cm}}] {
    Kicker & Receiver & $\gamma_{\text{1st half}|t}$ & $\gamma_{\text{2nd half}|t}$ \\
    Player (1) & Player (3) & \node[color=red]{\textbf{1.0}}; & \node[color=red] (a) {0.0}; \\
    Player (3) & Player (5) & \node[color=red]{\textbf{1.0}}; & \node[color=red]{0.0};  \\
    Player (3) & Goal   (5) & \node[color=red]{\textbf{1.0}}; & \node[color=red]{0.0};  \\
    Player (2) & Player (4) & \node[color=red]{\textbf{1.0}}; & \node[color=red]{0.0};  \\
    ... & ... & ... \\
    Player (4) & Goal   (2) & \node[color=blue]{0.0}; & \node[color=blue]{\textbf{1.0}};  \\
    Player (2) & Player (1) & \node[color=blue]{0.0}; & \node[color=blue]{\textbf{1.0}};  \\
};
\draw[] (pgthalves-2-1.north west) -- (a.north east);

\matrix (theta_offense)[matrix of math nodes,left delimiter={(},right delimiter={)}, above right = -0.6cm and .65cm of pgthalves, anchor = north west, inner sep=.5pt,outer sep=0, minimum height=0.4cm,  minimum width=0.4cm,
style={nodes={color=red}}
] {
	0 & 0 & \textbf{\nicefrac{3}{4}} & \textbf{\nicefrac{1}{4}} & 0 \\
	0 & 0 & \textbf{\nicefrac{1}{4}} & \textbf{\nicefrac{3}{4}} & 0 \\
	0 & 0 & 0 & 0 & \textbf{1} \\
	0 & 0 & 0 & 0 & \textbf{1} \\
};
\draw  [decoration={brace},decorate,line width=1pt] ([yshift=-0.2cm]theta_offense.south east) -- ([yshift=-0.2cm]theta_offense.south west) 
node (offense_label) [midway, below, yshift=-0.1cm,align=center] {$\mphi_\text{offense}$};

\matrix (theta_defense)[matrix of math nodes,left delimiter={(},right delimiter={)}, right = .5cm of theta_offense, inner sep=.5pt,outer sep=0, minimum height=0.4cm,  minimum width=0.4cm,
style={nodes={color=blue}}
] {
	0 & \textbf{1} & 0 & 0 & 0 \\
	\textbf{1} & 0 & 0 & 0 & 0 \\
	\textbf{\nicefrac{1}{4}} & 0 & 0 & \textbf{\nicefrac{3}{4}} & 0 \\
	0 & \textbf{\nicefrac{1}{4}} & \textbf{\nicefrac{3}{4}} & 0 & 0 \\
};
\draw [decoration={brace},decorate,line width=1pt] ([yshift=-0.2cm]theta_defense.south east) -- ([yshift=-0.2cm]theta_defense.south west) 
node (defense_label) [midway, below, yshift=-0.1cm, align=center] {$\mphi_\text{defense}$};

\begin{pgfonlayer}{bg}
    \node (bg1)[above left = .2cm and .5cm of p_gt_0, fill=gray!20, anchor=north west, minimum height=3.7cm, minimum width=11.7cm] {};
    \node (bg1_label) [left = -0.0cm of bg1, rotate=90, fill=gray!30, align=center, minimum width=3.7cm, anchor=center, align=center] {a) \textbf{homogeneous} hypothesis\\$H_\text{hom}=(\mgamma_\text{one}, (\mphi_\text{uniform}))$};
    
    \node (bg2) [above left = .2cm and .5cm of pgthalves, fill=gray!20, anchor=north west, minimum height=3.7cm, minimum width=11.7cm] {};
    \node (bg2_label) [left = -0.0cm of bg2, rotate=90, fill=gray!30, align=center, minimum width=3.7cm, anchor=center] {b) \textbf{heterogeneous} hypothesis\\$H_\text{het}=(\mgamma_\text{half}, (\mphi_\text{off.}, \mphi_\text{def.}))$};
    
    \node (box_pgt) [above left = 0.32cm and 0cm of p_gt_0, anchor = north west, minimum height = 7.9cm, minimum width = 5.3cm, dashed, draw=black, inner sep =0.0cm] {};
    \node (box_theta) [right = 0.15cm of box_pgt.north east, anchor = north west, minimum height = 7.9cm, minimum width = 5.6cm, dashed, draw=black, inner sep = 0.5cm] {};
    
    \node (label_pgt) [below = 0.0cm of box_pgt, anchor = north, align=center] {group assignment probabilities $\mgamma_{\cdot|t}$\\
    for each transition $t \in D$};
    \node (label_theta) [below = 0.0cm of box_theta, anchor = north, align=center] {transition probabilities $\mphi_g$\\for each group $g \in G$};

\end{pgfonlayer}

\end{tikzpicture}
     \caption{\textbf{Hypotheses for heterogeneous sequence data.}
    In \approachname{}, we formulate hypotheses about heterogeneous sequence data.
    E.g., in the soccer example, we define two hypotheses:
    The homogeneous hypothesis $H_\text{hom}$ (a) assumes that players just randomly pass the ball around; the heterogeneous hypothesis $H_\text{het}$ (b) assumes an offensive strategy in the first half of the game and a defensive strategy in the second half, cf. \Cref{fig:figure1}.
    This is formalized based on two components: \emph{group assignment probabilities} $\mgamma$, i.e., probability distributions over a set of groups for each transition, and a belief matrix of \textit{group transition probabilities} $\mphi_g$ for each group $g$. 
    The soccer example features a special case, where group assignments are deterministic, i.e., the probabilities are either 0 or 1.
    }
    \label{fig:hypotheses}
\end{figure}

\para{Example}
For illustration, consider again the soccer game example from~\Cref{fig:figure1}.
In the following, we specify two hypotheses for this scenario: a homogeneous one $H_\text{hom}$ and a heterogeneous one $H_\text{het}$.
The homogeneous hypothesis $H_\text{hom}$ expresses the belief that the players just kick around randomly.
This can be formalized as a single matrix of transition probabilities $\mphi_\text{uniform}$ as shown in~\Cref{fig:hypotheses}(a).
Consequently, the corresponding group assignment probabilities $\mgamma_\text{one}$ only assign transitions to a single group. 
As a more fine-granular hypothesis using a heterogeneous structure, $H_\text{het}$  assumes that the soccer team played by an offensive strategy in the first half of the game and by a defensive strategy in the second half.
For this, we need two separate transition probability matrices ($\mphi_\text{offense}$ and $\mphi_\text{defense}$), one for each halftime. 
Then, we assign each transition to the group (halftime) it belongs to via $\mgamma_\text{half}$.
Transitions are assigned to half-times without uncertainty, thus, the probabilities used are either 0 or 1.
The resulting hypothesis is defined as $H_\text{het} = (\mgamma_\text{half}, (\mphi_\text{offense}, \mphi_\text{defense}))$ as visualized in~\Cref{fig:hypotheses}(b).
Now, our approach \approachname{} determines the marginal likelihood $\Pr(D|H_\text{hom})$ and $\Pr(D|H_\text{het})$ as a measure for the plausibility of the data under a hypothesis.
Since $\Pr(D|H_\text{het}) > \Pr(D|H_\text{hom})$ (as demonstrated later, \Cref{sec:example}), we assert that explaining the data as a result of an offensive strategy in the first half of the game and a defensive strategy in the second half ($H_\text{het}$) is a more plausible hypothesis given the observed data.

\para{Flexibility}
The soccer example from above features an important special case of our approach, i.e., for the heterogeneous hypothesis, the assignment of transitions to groups is \emph{deterministic} $\mgamma_{g|t} \in \{0,1\}$.
However, our method also supports arbitrary group assignment probabilities. 
This is can be useful when hypotheses assume gradual change between generating processes (e.g., the team continuously switches from offense to defense during a game), when they suggest that the generating entity switches between different processes (e.g., when the team unpredictably switches between offensive and defensive play)\mbe{... mich interessiert aber echt immernoch der zusammenhang zu hypothesen, die man direkt mischt ...}, or if there is uncertain or insufficient information available (e.g., the time of some passes was not accurately recorded).

\begin{revised}
Overall, the ability to specify group assignment probabilities allows to formulate very intricate dependency structures and may serve as an interface to more complex, possibly latent processes.
In particular, group assignment probabilities and consequently the transition probabilities associated with each transition can depend on any information associated with a transition, specifically including background information (e.g., user properties, length and duration of the sequence, state properties, time of the day), information derived from previously as well as subsequently visited states, or even information about other traces.
For instance, this allows for hypotheses modelling higher order Markovian processes, i.e., by defining $n^x$ groups (where $n$ is the number of states and $x$ is the order of the model) and setting the group assignment probabilities depending on the state history of each transition. 
Some concrete examples on defining hypotheses that take into account the overall sequence are featured in the experimental evaluation in \Cref{sec:experiments}.Thus, even though there are some limitations and possible extensions (cf. \Cref{sec:discussion}), all in all, \approachname{} provides a very flexible and easy to use framework to model a very large and possibly complex set of hypotheses.
\end{revised}

\subsection{The Mixed Transition Markov Chain (MTMC) Model}
\label{sec:model}
\vspace{-2mm}
A standard Markov chain model is unable to capture heterogeneity in sequential data.
Therefore, we propose the \emph{Mixed Transitions Markov Chain} (MTMC) model as an extension for which we can formulate heterogeneous hypotheses as beliefs over its parameters.

MTMC assigns each transition $t \in D$ in the dataset to a group $g \in G = \{g_1, ..., g_o\}$, which is drawn from an individual categorical distribution with parameters $\vgamma_{t} = (\mgamma_{g_1|t}, \ldots, \mgamma_{g_o|t})$, where $\mgamma_{g|t}$ denotes the probability of transition $t$ belonging to group $g$. 
Then, given a common state space, each group $g \in G$ is associated with its own first-order Markov chain.
Thus, for each source state $s_i$, there is a categorical distribution $\vtheta_{s_i|g} = (\theta_{i,1|g}, \ldots, \theta_{i,n|g})$ over all potential target states.
The parameters $\theta_{i,j|g}$ 
are distributed according to a (prior) Dirichlet distribution $Dir(\valpha_{s_i|g})$ with hyperparameters $\valpha_{s_i|g} = (\alpha_{i,1|g}, \ldots, \alpha_{i,n|g})$.
For shorter notation, we write the set of transition probabilities over all groups as $\mtheta = (\mtheta_1, \ldots, \mtheta_o)$ and the set of transition probabilities over all states in a group as $\mtheta_g = (\vtheta_{s_1|g}, \ldots, \vtheta_{s_n|g})$.
Similarly, we denote the set of all hyperparameters for all Markov models, i.e., all Dirichlet parameters, as $\malpha = (\malpha_1, \ldots, \malpha_o)$, and the set of all hyperparameters for a single group as $\malpha_g = (\valpha_{s_1|g}, \ldots, \valpha_{s_n|g})$.
Finally, we write the set of all group assignment probabilities for all transitions in the dataset as $\mgamma = (\vgamma_t)$ with $t \in D$.
Given these definitions, considering only a single group ($|G| = 1$), MTMC is a direct generalization of the a first-order Markov chain model.

Overall, the MTMC model is described by the following generative process that, given a set of transitions $D = \{t_1, \ldots, t_m\}$, generates for each transition $t_k \in D$, a destination state $\text{dst}_k$ for a known source state $\text{src}_k$ and known group assignment probabilities $\vgamma_{t_k}$:
\begin{enumerate}
    \item For each group $g \in G$ and each state $s_i \in S$, \\
    choose transition probabilities $\mtheta_{s_i|g} \sim Dir(\malpha_{s_i|g})$.
    \item For each transition $t_k$:
    \begin{enumerate}
        \item Choose the group assignment $z_k \sim Cat(\vgamma_{t_k})$.
        \item Choose the destination state $\text{dst}_k \sim Cat(\vtheta_{\text{src}_k|z_k})$.
    \end{enumerate}
\end{enumerate}

\subsection{Eliciting priors from hypotheses}
\label{sec:eliciting}
As mentioned in \Cref{sec:problem}, \approachname{} elicits hypotheses as Bayesian priors for the MTMC model (see \Cref{sec:model}), which takes two independent sets of parameters: 
the group assignment probabilities $\mgamma$ and 
the prior parameters $\malpha$.
While the group assignment probabilities are directly specified by a hypotheses $H = (\mgamma, \mphi)$, see \Cref{sec:problem}, the parameters $\malpha$ of the Dirichlet prior need to be \emph{elicited} from the transition probabilities $\mphi$ defined by the hypothesis.

\para{Deterministic Assignments. }
For \emph{deterministic} group assignments, i.e., $\gamma_{g|t} \in \{0,1\}$, we determine the parameters $\malpha_g$ of the Dirichlet distributions for each group $g \in G$ separately, using the notion of \emph{pseudo-observations}, cf.~\cite{singer2015}.
That is, for each group $g \in G$ and each state $s_i$, we set the Dirichlet parameters starting from an uninformed proto-prior and add $\kappa$ transitions distributed as the hypothesis suggests for this group via $\mphi_g$.
Formally, this is:
\begin{align}
\alpha_{i,j|g} = \kappa \cdot \phi_{i,j|g} + 1.
\label{eq:prior_deterministic}
\end{align}
Here, the number of pseudo-observations $\kappa$ (also called concentration factor) reflects the strength of belief in the respective hypothesis. Different settings for the concentration parameter lead to different priors. In our approach, we compare hypotheses along a range of different concentration factors, i.e., strengths of belief in the respective hypothesis.

For example, consider the heterogeneous hypothesis $H_\text{het} = (\mgamma_\text{half}, (\mphi_\text{offense}, \mphi_\text{defense}))$ from \Cref{fig:hypotheses}(b). 
It features two groups (the first and second half of a soccer game), and for each group $g \in \{\text{1st half}, \text{2nd half}\}$ it defines specific beliefs in certain transition probabilities, via the matrix entries $\phi_{i,j|g}$.
For each group, a matrix of prior parameters $\malpha_g$ is determined according to~\Cref{eq:prior_deterministic}.
The offense hypothesis for the first half suggests transition probabilities $\mphi_{s_1|\text{1st half}}$\small$=(0, 0, \nicefrac{3}{4},  \nicefrac{1}{4}, 0)$\normalsize{} for the first row of the transition probability matrix.
Choosing an arbitrary concentration factor of $\kappa = 10$, we therefore obtain a Dirichlet prior with parameters $\malpha_{s_1|\text{1st half}}$\small$= (1, 1, 8.5,  3.5, 1)$\normalsize.

\para{Probabilistic Assignments. }
For \emph{probabilistic} group assignments, i.e., $0 < \gamma_{g|t} < 1$, we need to adapt these basic priors to account for misassignments of groups.
For example, consider a scenario in which the dataset is divided into two groups that behave completely different.
Then, if some transitions cannot be assigned to groups with certainty, the model will randomly associate some transitions which behave like the first group with the second group, and vice versa.
Thus, given uncertain group assignments, the behavior expected from a set of transitions assigned to one group is actually a mixture of behavioral traits of both groups.
Consequently, we compute the number of  pseudo-observations of the Dirichlet priors for a group $g$ as a mixture of hypotheses that is determined by the group assignment probabilities of all transitions.
For that purpose, for each transition $t_k$, we compute the probability that the model assigns $t_k$ to group $g$ although it actually belongs to group $g'$ ($\gamma_{g|t_k} \cdot \gamma_{g'|t_k}$).
This probability is then used as a weight for the respective belief matrix $\mphi_{g'}$.
Formally:
\begin{align}
\alpha_{i,j|g} = \kappa \cdot \left( \frac{1}{Z_i} \cdot \sum_{t_k \in D} \left( \sum_{g' \in G} \gamma_{g|t_k} \cdot \gamma_{g'|t_k} \cdot \phi_{i,j|g'} \right)\right) + 1 \text{,}
\end{align}
where $\nicefrac{1}{Z_i}$ represents a normalization factor to ensure that the transition probabilities from each state to the other states in the mixture sum up to 1.
Note that for deterministic group assignments, the formula simplifies to~\Cref{eq:prior_deterministic}.

\fle{Add example here? If yes, which one? creating data is ugly because we need a set of transitions}

\subsection{Model Inference}
\label{sec:inference}

For comparing the plausibility of heterogeneous hypotheses, in \approachname{}, we determine the evidence (marginal likelihood) of the data under a hypothesis (cf. \Cref{sec:problem}) based on the MTMC model as introduced in \Cref{sec:model}.
The marginal likelihood can be understood as an average over the likelihood of all parameter settings weighted by their prior probability (given by the hypothesis). This can be written as an integral over all parameter settings $\mtheta$:  
\begin{align*}
\Pr(D|H) 
& = \numberthis\label{eq:ml_int} \int\underbrace{\Pr(D | \mtheta,\mgamma)}_{\text{likelihood}} ~ \underbrace{\Pr(\mtheta|\malpha)}_{\text{prior}} ~ d\mtheta 
\end{align*}

In the remainder of this section, we elaborate on how to compute the marginal likelihood for our MTMC model given some observed data and any hypothesis.
We start by deriving an analytical solution.
However, the resulting formula is computationally intractable for non-trivial datasets.
Thus, we show that for the special case of hypotheses with deterministic group assignments, the calculation can be substantially simplified. 
Additionally, for the general case, we explain how it can be efficiently
approximated by sampling.

\para{Analytical solution. }
\label{sec:method-analytical}
When ignoring the group assignment probabilities $\mgamma$ in \Cref{eq:ml_int}, the marginal likelihood of the MTMC model is equivalent to the homogeneous Markov chain model for which an analytical solution exists~\cite{singer2014detecting}. However, in our setting, we need to aggregate  over all possible instantiations $\omega \in \Omega$ of group assignments. 
Each instantiation $\omega$ maps each transition $t$ to a group $\omega(t)$. 
The probability $p_\omega$ of an instantiation $\omega$ is determined by the group assignment probabilities specified in the hypothesis, i.e., $p_\omega = \prod_{t \in D} \gamma_{\omega(t)|t}$.
For a fixed assignment to groups, we can then determine the overall marginal likelihood as the product of marginal likelihoods of the individual groups. For each group, the marginal likelihood can be calculated analytically as a combination of beta functions over the hyperparameters for that group, and over the observed counts in the data according to the fixed group assignment (see~\cite{singer2014detecting} for details).
Overall, we obtain the following formula (for an in-depth derivation see \Cref{app:ml}):

\begin{align*}
\Pr(D|H) 
&= \numberthis\label{eq:ml-analytical} \sum_{\omega \in \Omega} p_\omega \prod_{g \in G} \prod_{s_i \in S} \frac{ B(\vn_{s_i|g,\omega} + \valpha_{s_i|g})}{B(\valpha_{s_i|g})} \text{,}
\end{align*}

Thus, the marginal likelihood of MTMC can be seen as a weighted average over the marginal likelihood of all possible group assignments $\omega$.
Unfortunately, this solution is computationally intractable for real world datasets because the number of different group assignments $|\Omega|$ grows exponentially with each additional \emph{transition} $t \in D$.

However, we can substantially decrease the computational costs for the important special case of deterministic group assignments, i.e., where the group assignment probabilities are either zero or one.
Then, there is only one valid instantiation of the group assignments, i.e., all but one weight $p_\omega$ are zero, and the formula from \Cref{eq:ml-analytical} simplifies to:
\begin{align*}
\Pr(D|H) 
&= \prod_{g \in G} \prod_{s_i \in S} \frac{ B(\vn_{s_i|g} + \valpha_{s_i|g})}{B(\valpha_{s_i|g})}
\end{align*}

Thus, in this case, the marginal likelihood is equivalent to the product over the marginal likelihoods across all groups.
This can be calculated much more efficiently as the computational complexity only linearly depends on the number of states and groups.
The formula also allows for leveraging existing parallelized approaches like SparkTrails \cite{becker2016sparktrails}.

\para{Approximation. }
\label{sec:method-approximation}
For the general, probabilistic case, calculating the marginal likelihood of an MTMC model analytically with ~\Cref{eq:ml-analytical}  is  computationally intractable.
Therefore, we show how we can efficiently approximate it by direct sampling.
According to the formula, the overall marginal likelihood is a weighted average over the marginal likelihoods of all group assignments $\Omega$.
To approximate this, we sample from the space of all group assignments $\Omega$ according to their respective probability $p_\omega$ and calculate the average marginal likelihood given these sampled group assignments $\Pr(D|\malpha,\omega)$.
Since for individual transitions the process of choosing groups is independent from each other, a single group assignment can be sampled by drawing the group $z_k$ for each transition $t_k \in D$ according to its group assignment distribution $z_k \sim Cat(\vgamma_{t_k})$ (also see the generative process in \Cref{sec:model}).
The sampling procedure follows the intuition that factors with small group assignment probabilities contribute less to the overall marginal likelihood.
Formally, we can compute the approximated marginal likelihood from a list of sampled group assignments $\Omega'$ as:
\begin{align*}
\Pr(D|H)
&\approx \frac{1}{|\Omega'|} \sum_{\omega \in \Omega'} \underbrace{\prod_{g \in G} \prod_{s_i \in S} \frac{ B(\vn_{s_i|g,\omega} + \valpha_{s_i|g})}{B(\valpha_{s_i|g})}}_{\Pr(D|\malpha,\omega)} 
\end{align*}

In our experiments, we found that the results are stable for very small numbers of iterations (less than $50$) if the number of transitions is sufficiently high.
This allows to run our experiments in \Cref{sec:experiments} in only a few hours on a regular desktop machine.

\subsection{Visualizing and interpreting results}
\label{sec:example}
\vspace{-1mm}

In this section, we describe our recommended way of performing experiments, visualizing results, and interpreting them.
To this end we use the soccer example from \Cref{fig:figure1} and investigate which strategies the soccer team has used. 
For instance, they may have passed the ball randomly, or they may have played by a more intricate strategy. 
More specifically, given the observed transitions from \Cref{fig:figure1}(a-c), we aim to compare the plausibility of the different beliefs in transition probabilities from \Cref{fig:figure1}(d-g) utilizing the marginal likelihood as elaborated in \Cref{sec:inference}. 
In particular, we study the four hypotheses \emph{uniform}, \emph{offense}, \emph{left-flank}, and \emph{defense}, as well as a \emph{data} hypothesis.
The latter uses the actual observed transition probabilities as belief; thus it is only used for comparison.
We consider these beliefs for three group assignments: 
(a) a homogeneous one (all transitions are in one group), 
(b) a group assignment defined by the half-time of the passes/shots, and 
(c) a completely random group assignment.
The hypotheses are formulated analogously to the examples covered in \Cref{sec:eliciting}.
\begin{figure}[t!]
    \vspace{-.8em}
    \centering
    \subfigure {
        \includegraphics[scale=0.5]{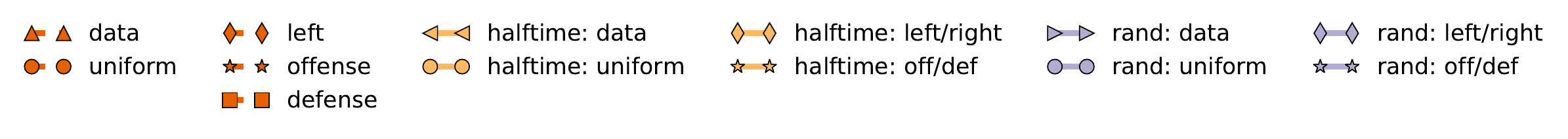}
    }
    \addtocounter{subfigure}{-1}
    \subfigure[Homogeneous hypotheses] {
        \label{fig:toy_homo}
        \includegraphics[scale=0.5]{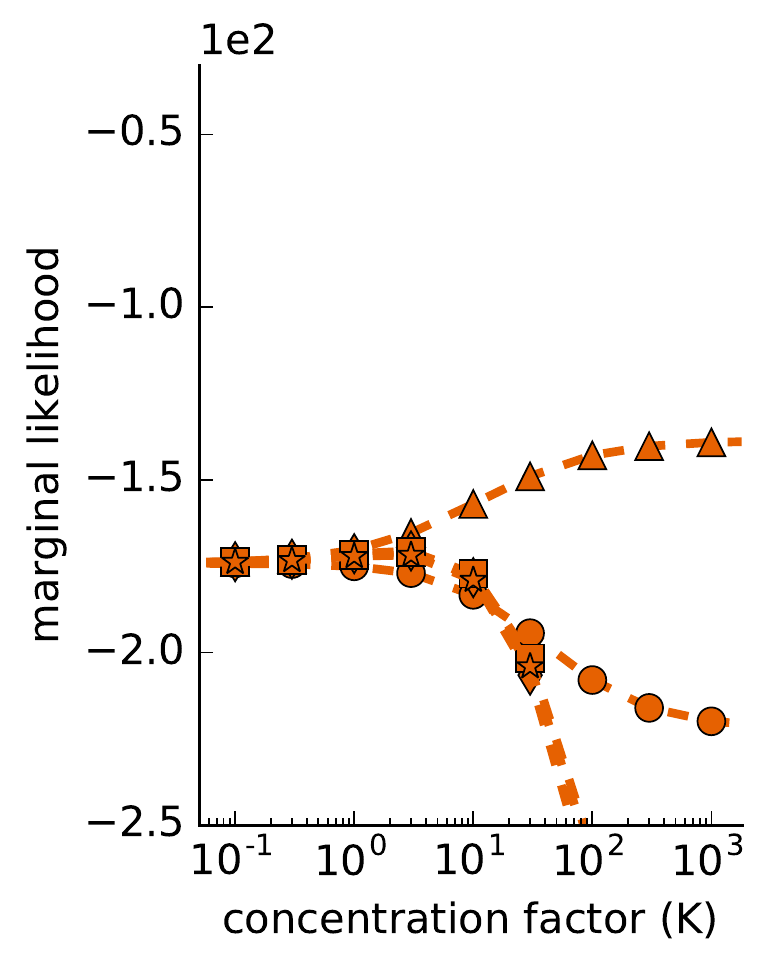}
    }
    \subfigure[Heterogeneous hyp. based on a split according to halftimes] {
        \label{fig:toy_half}
        \includegraphics[scale=0.5]{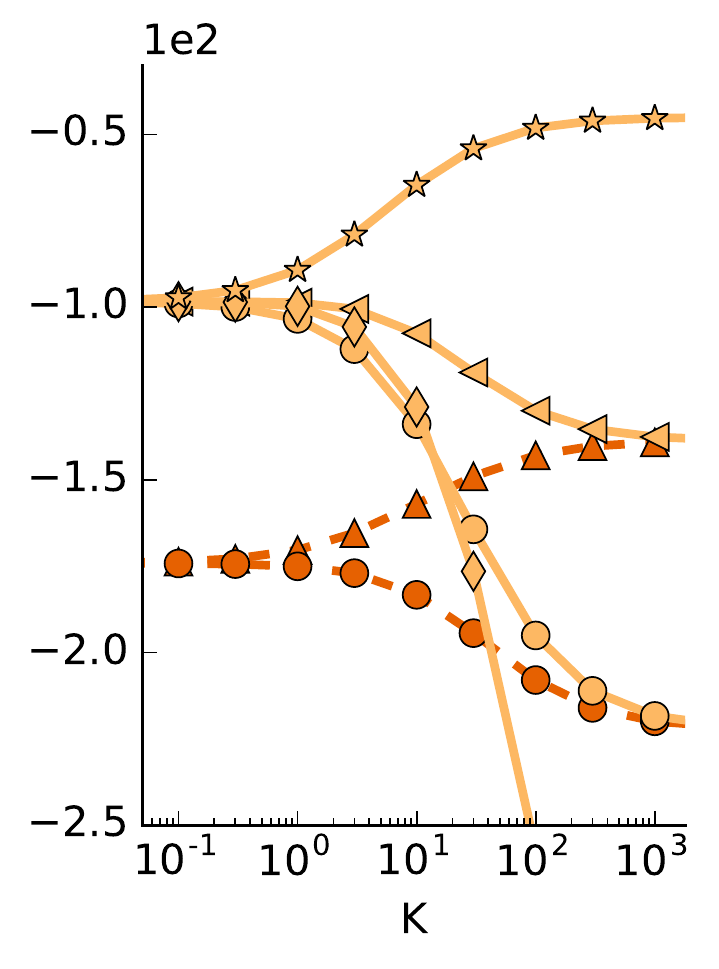}
    }
    \subfigure[Heterogeneous hypotheses based on a random split] {
        \label{fig:toy_random}
        \includegraphics[scale=0.5]{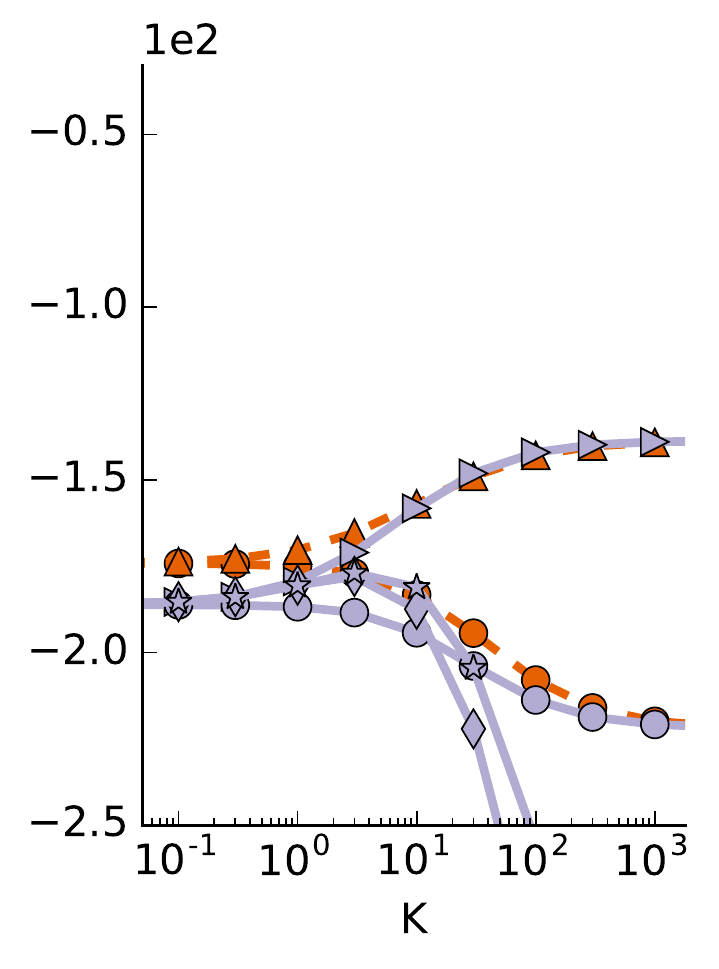}
    }    \caption{\textbf{Results for the illustrating example. }
    This plot shows the \approachname{} results for the illustrating soccer example, i.e., marginal likelihood values of different hypotheses for increasing strengths of belief $\kappa$.
    We observe that among the hypotheses without grouping, the uniform hypothesis performs best (a). However, far more plausible explanations can be obtained by heterogeneous hypotheses that assume different behavior in both halftimes (b). Finally, randomly splitting the data into groups leads to less plausible explanations (c).
    }
    \vspace{-1em}
    \label{fig:toy}
\end{figure}
The results are shown in \Cref{fig:toy}(a-c).
In each plot, the x-axis denotes increasing values of the concentration factor $\kappa$, which expresses an increasingly strong belief in the hypotheses. The y-axis shows the marginal likelihood; each line represents one given hypothesis; solid lines refer to heterogeneous hypotheses and dashed lines to homogeneous hypotheses. In general, higher values of the marginal likelihood indicate more plausible hypotheses. 

\begin{revised}
\para{Relativity. }
An essential issue for interpreting the results from MixedTrails (or any method using Bayes factors) is that results are \emph{relative}.
This means that even if one hypothesis outperforms all other hypotheses under consideration, this does not necessarily mean that it models the data well.
However, the goal of this paper is to compare existing hypotheses from literature, domain experts, ideas, or intuition. 
The goal is not to find models which perform well for prediction or similar tasks.
Nevertheless it may be desirable to validate the hypotheses with regard to their generative quality.
For this, we suggest the comparison with the uniform hypothesis (as we do in this example) or a with a hypothesis with a flat (uninformed) prior ($\kappa = 0$).
The former assumes all \emph{transitions} to be equally likely, while the latter is equivalent to assuming that all transition probability \emph{distributions} are equally likely.
Also, additional baselines can arise naturally in specific application domains. 
For example, if analyzing navigation behavior between web pages, a baseline could be that only transitions to \emph{linked} pages are equally likely, and not to all web pages in the dataset (cf.~\cite{dimitrov2016}).
We consider the relative order of hypotheses as still viable and interesting if the hypotheses are better than such a baseline hypothesis because they cover at least some aspects of the transition processes.
At the same time, if all hypotheses perform worse than the flat prior ($\kappa = 0$), then the data may be too complex for the chosen hypotheses, or the facilitated background data is not sufficient to explain the underlying processes.

\para{Significance. }
With regard to the significance of differences, we refer to Kass and Raftery's established interpretation table \cite{kass1995bayes}.
This means that conclusions should only be drawn for sections of the marginal likelihood plots where the values are farther apart than 10. 
In these cases, the change of the posterior is to be interpreted as ``decisive''.
Consequently, we only draw conclusions from such decisive results in this manuscript.
\end{revised}

\para{General properties of curves. }
Different values along the x-axis enable interpretation beyond providing a relative order of hypotheses: 
For the left-hand side of the plots (values of $\kappa$ close to zero) the influence of the transition probabilities of a hypothesis is very weak and the marginal likelihood depends mostly on the group assignment. 
\begin{revised}
Thus, the higher the marginal likelihood for $\kappa = 0$, the more a heterogeneous hypothesis can benefit if it models the transition probabilities in each group correctly.

For growing values of $\kappa$, the Bayesian framework increasingly takes into account the quality of the chosen transition probabilities for the corresponding group assignments.
At first it allows for a large tolerance, i.e., it integrates over variations of the specified transition probabilities. 
Then, it consecutively decreases this tolerance, requiring that the transition probabilities are very precise.
For very high values of $\kappa$, the marginal likelihood converges towards the likelihood of the hypothesis.
Consequently, the marginal likelihood of heterogeneous hypotheses that assume identical transition behavior in all groups converges towards their homogeneous counterparts (cf. \emph{uniform} and \emph{1st/2nd: uniform} in \Cref{fig:toy_half}).
This is because there is no difference between a homogeneous and a heterogeneous hypothesis if the transition probabilities in each group describe the same generative process.

Overall, the relation of hypotheses along increasing concentration factors gives intricate information about the influence of the different components of the hypotheses.
\end{revised}

\para{Results on homogeneous hypotheses. }
\Cref{fig:toy_homo} shows results for the homogeneous hypotheses.
As expected, the data ``hypothesis'', which uses the actual observed transitions, achieves the highest marginal likelihood values for all $\kappa$.
Apart from that, the uniform hypothesis explains the observed transitions best.
The left-flank, the offense, and the defense hypothesis exhibit strongly decreasing marginal likelihoods for an increasing concentration factor, which indicates that these hypotheses are not supported by the observed data.
These results can analogously be obtained by the HypTrails approach~\cite{singer2015}.

\para{Results on heterogeneous hypotheses: the split. }
However, our approach \approachname{} enables us to also compare more fine-grained, \emph{heterogeneous} hypotheses.
\Cref{fig:toy_half} features four heterogeneous hypotheses (solid lines) that assign the data deterministically into two groups, i.e., the first and the second half-time. Additionally, it shows the homogeneous data hypothesis and the uniform hypothesis for comparison (dashed lines).
For a concentration factor $\kappa = 0$ the marginal likelihood depends only on the group assignment.
\begin{revised}
Therefore hypotheses with the same group assignment probabilities start at the same marginal likelihood level. 
\end{revised}
Now, since our dataset indeed features different behavior in both halftimes as the group assignment of our heterogeneous hypotheses suggests, their marginal likelihood is higher compared to the homogeneous hypotheses at $\kappa = 0$. This indicates how strongly the split divides transitions into differing processes, before delving deeper into the plausibility of the expressed hypotheses with an increasing concentration factor $\kappa$.

\para{Results on heterogeneous hypotheses: the curve. }
For higher values of $\kappa$, the marginal likelihoods diverge:
The offense/defense hypothesis, i.e., in the first half-time players behave as the offense belief suggests, and in the second halftime as the defense belief suggests (see \Cref{fig:hypotheses}), is fully supported by the observed data and thus yields the highest values for all $\kappa$.
In comparison to the homogeneous hypotheses, this curve can be interpreted as:
\emph{``This hypothesis features a good group assignment and the transition beliefs reflect the behavior in the observed data better.''}
If we assign the same belief in transition probabilities to both halftimes, e.g., uniform probabilities, or the globally observed transition probabilities (data), then smaller values are obtained, indicating that these transition beliefs differ from the observed data.
Additionally, for very large values of $\kappa$, the scores converge with the ones from the respective homogeneous hypothesis because the corresponding heterogeneous hypothesis does not define different transition probabilities for each group, which eventually nullifies the effect of the split.
Finally, if we use transition beliefs that are not actually supported by the data for both groups, e.g., a left-flank and right-flank preference in the two halftimes, then the marginal likelihood curve rapidly declines.
The respective curve can --- in comparison to the other curves --- be interpreted as:
\emph{``The hypothesis uses a good group assignment, but the transition beliefs are not reflected in observed data.''}

\para{Results for a random split and summary. }
\Cref{fig:toy_random} shows the same four hypotheses, but assigns transition to two groups randomly (\emph{rand}).
Since a random group assignment increases the model complexity, but does not allow for a better model of transition behavior, all hypotheses start with a lower value than the homogeneous hypotheses on the left hand side of the plot. For larger values of $\kappa$, we can see the same convergence behavior as before, but, overall, the marginal likelihoods of the heterogeneous hypotheses are substantially lower and also rank lower than their homogeneous counterparts.
Overall, these examples give a broad overview of possible \approachname{} results.
More examples are covered in \Cref{sec:experiments}.

% \mbe{a) somehow I would ask: then why not use the right hand-side? and b) I am very close to suggesting to cartesian grouping again ... because it just jumps at you when lookng at the graph (uniform/uniform) :D it may be explainable in a few very short sentences ... however, what would be a good real world scenario to use it? c) should we try to go further in interpreting plots? when high at first but low later ... blubb}
\section{Experiments}
\label{sec:experiments}

In this section, we demonstrate the applicability and benefits of our approach with experiments on synthetic and real-world datasets.
\begin{revised}
An open source Python implementation\furl{http://dmir.org/mixedtrails} as well as the datasets are freely  available\footnote{\revised The scripts for generating the synthetic data are included in the code, the Wikispeedia data set (cf. \Cref{sec:wiki}) is accessible online and the Flickr data (cf. \Cref{sec:flickr}) is available via e-mail to Martin Becker.}.
Conclusions from the experimental results drawn in the text rely on results that are ``decisive'' with respect to the established interpretation table given in \cite{kass1995bayes}, cf. \Cref{sec:example}.
\end{revised}

\vspace{-5mm}
\subsection{Synthetic Datasets: Deterministic Group Assignments}
\label{sec:synth}
\vspace{-2mm}
First,  we consider three synthetic examples in order to showcase the properties of \approachname{} in a controlled setting.
For each example, we generate a transition dataset according to a predefined mechanism and compare the plausibility of several homogeneous and heterogeneous hypotheses. We show that those hypotheses that best capture the known mechanism generating the synthetic data are indeed reported as the most plausible ones.

\para{Datasets. }
\label{sec:synth_data}
The synthetic transition datasets are based on a random  Barab\'{a}si-Albert preferential attachment graph~\cite{Barabasi1999} with 100 nodes and 10 edges for each new node.
Each node has a random color $c \in \{\text{red}, \text{blue}\}$ assigned with a probability of $p_c = 0.5$.
From this graph, we derive three different transition datasets generated by 10,000 random walkers with different characteristics.
Just like each state, each walker also has a color $c \in \{\text{red}, \text{blue}\}$ assigned randomly with $p_c = 0.5$.
Each walker chooses her first node randomly and navigates through the network generating transitions depending on the mechanism for the respective dataset. She stops after ten steps.
For the first dataset $D_\text{link}$, we consider \emph{link} walkers that choose the next node uniformly from all adjacent nodes, independent of the walker color.
This corresponds to a transition probability matrix $\mtheta_\text{link}$ equal to the (row-wise) normalized adjacency matrix of the underlying graph.
For the second dataset $D_\text{color}$, walkers of the ``red'' (``blue'', respectively) group exclusively transition according to a probability matrix $\mtheta_\text{red}$ ($\mtheta_\text{blue}$) which adapts $\mtheta_\text{link}$ such that transitions to red (blue) nodes are ten times more likely. 
The third dataset $D_\text{mem}$ is generated by ``memory walkers'' that dynamically choose their next state based on their history, i.e., they use a different transition matrix dependent on the colors of the states they have already visited (including the current state).
In particular, if they have visited more red than blue nodes, they use the matrix $\mtheta_\text{red}$, and if they have visited more blue than red nodes, they use the matrix $\mtheta_\text{blue}$.
In case of a draw, they use the random transition matrix $\mtheta_\text{link}$.

\para{Hypotheses. }
\label{sec:synth_hypo}
For the three datasets we construct corresponding hypotheses:
first, the homogeneous hypothesis $H_\text{link} = (\mgamma_\text{link}, \mphi_{link})$, which expresses the belief that all transition are randomly chosen from the link network, thus $\mphi_{link} = (\mtheta_{link})$;
secondly, the color-preference hypothesis $H_\text{color} = (\mgamma_\text{color}, \mphi_\text{color})$ maps each transition to a group based on the color assigned to its walker and uses the actual probability matrices for the transitions in the groups as belief matrices: $\mphi_\text{color} = (\mtheta_\text{red}, \mtheta_\text{blue})$;
and thirdly, the memory hypothesis $H_\text{mem} = (\mgamma_\text{mem}, \mphi_\text{mem})$ reflects the generating mechanism in the third dataset:
The transitions are assigned to groups according to the majority of node colors already visited, and the transition belief matrix is constructed as described in the generation of the third dataset: $\mphi_\text{mem} = (\mtheta_\text{red}, \mtheta_\text{blue}, \mtheta_\text{link})$.
To illustrate how our approach copes with groups that introduce unnecessary complexity, we add a fourth hypothesis $H_\text{link-color} = (\mgamma_\text{color}, (\mtheta_{link}, \mtheta_{link}))$ that uses the grouping into ``red'' and ``blue'' walkers, but assumes the same movement behavior for both groups, i.e., equal transition likelihood for all links.

\para{Results. }
\label{sec:synth_results}
\begin{figure}[t!]
    \centering
    \subfigure{
        \label{fig:syn_legend}
        \hspace{2cm}
        \includegraphics[scale=0.5]{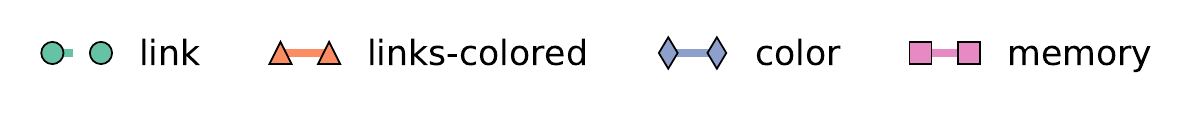}
        \hspace{2cm}
    }
    \addtocounter{subfigure}{-1}
    \subfigure[Link dataset] {
        \label{fig:syn_link}
        \includegraphics[scale=0.5]{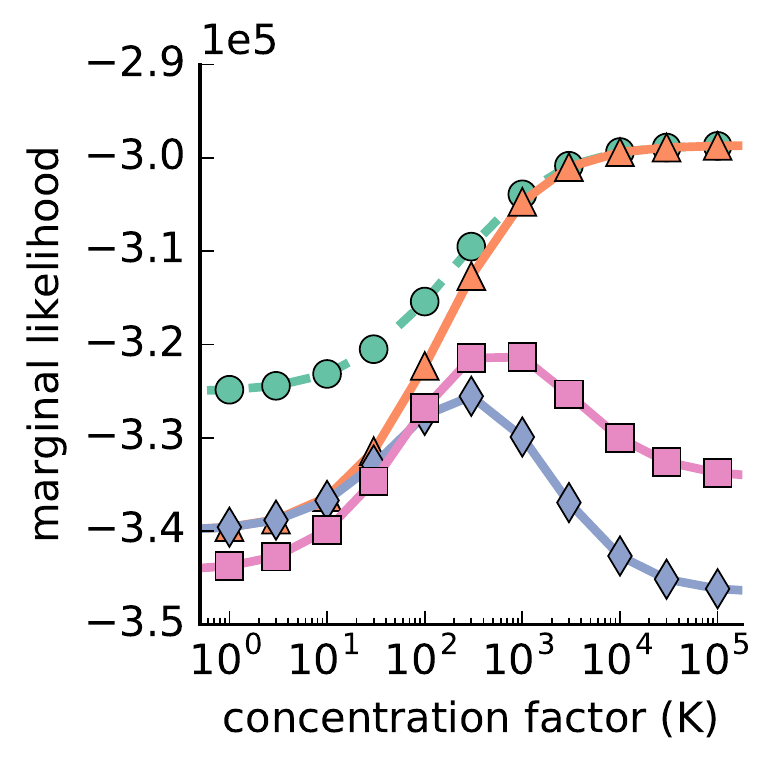}
    }
    \subfigure[Homophily dataset] {
        \label{fig:syn_homo}
        \includegraphics[scale=0.5]{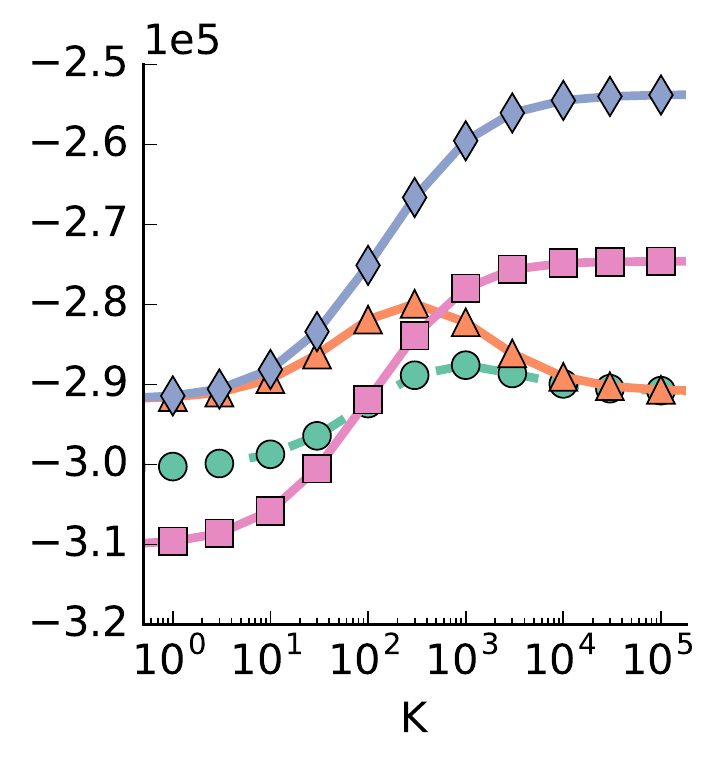}
    }
    \subfigure[Memory dataset] {
        \label{fig:syn_memory}
        \includegraphics[scale=0.5]{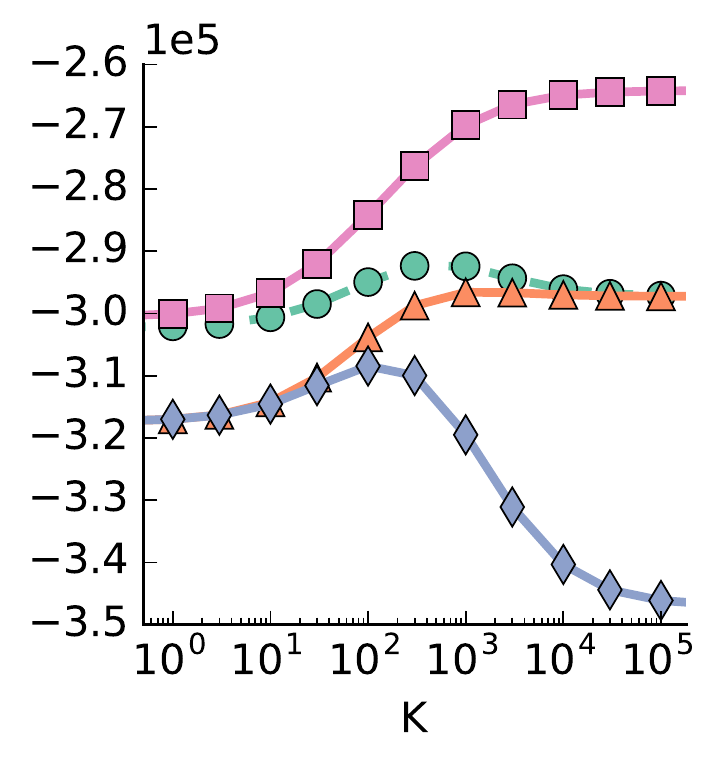}
    }
    \vspace{-2mm}
    \caption{\textbf{Synthetic data results.}
    We compare homogeneous ($H_\text{link}$) and heterogeneous hypotheses ($H_\text{link-colored}$, $H_\text{color}$ and $H_\text{mem}$) on three synthetic datasets ($D_\text{link}$, $D_\text{color}$ and $D_\text{mem}$).
    We observe that the hypotheses that are fitting the respective datasets work best, illustrating that the \approachname{} approach can identify the correct ordering of the defined hypotheses. 
    For details on interpreting the plots, see \Cref{sec:synth}.}
    \label{fig:synthetic}
    \vspace{-2em}
\end{figure}
Using \approachname{}, we compare these four hypotheses on all three datasets.
The results are visualized in \Cref{fig:synthetic}.
For the link dataset $D_{link}$, see~\Cref{fig:syn_link}, we find that the homogeneous hypothesis reflects the data very well and thus achieves the highest marginal likelihood (ML) values for all concentration factors.
The differences for small concentration factors $\kappa$ (left-hand side of the plot) indicate that the other group assignment probabilities used by the heterogeneous hypotheses do not introduce valuable information.
Both heterogeneous hypotheses show increasing ML for increasing $\kappa$ at first since the hypotheses carry some information, i.e., which network links are contained in the data.
With increasing concentration, however, the emphasis on some specific links (i.e., to red or to blue nodes), which is not reflected in the data, leads to a drop of the ML.
Furthermore, the memory hypothesis is closer to the data than the color hypothesis as it includes transitions to red and blue nodes in more equal proportions for each source state. 

Next, we consider the color dataset $D_\text{color}$, see~\Cref{fig:syn_homo}.
The ordering of the hypotheses on the left hand side of the plot indicates that
the assignment of transition into groups (by walker color) adds valid information to the corresponding hypotheses.
However, while the color preference hypothesis $H_\text{color}$ models the transition behavior within the groups very well, the grouped link hypothesis $H_\text{link-colored}$ does not.
This explains the diverging ML values for an increasing concentration factor.
When comparing the simple link hypothesis $H_\text{link}$ and the memory hypothesis $H_\text{mem}$, we observe that by introducing an incorrect grouping, the memory hypothesis  starts at a lower ML than the link hypothesis which does not introduce any groups.
However, with increasing concentration factors, the memory hypothesis starts to perform better, since, in contrast to the link hypothesis, it does incorporate the red and blue transition behavior even if on differing (but somewhat color-consistent) transition groupings.
Thus, overall, our model allows to establish a correct ordering on the given hypotheses based on the processes used to generate the data. 

Finally, we consider the memory dataset $D_{mem}$.
Here we can observe that --- as expected --- the memory hypothesis $H_\text{mem}$ performs best for all values of $\kappa$.
The group assignment according to walker colors does not correlate with the actual groups in the data and thus leads to lower ML value for low values of $\kappa$ compared to a homogeneous hypothesis.
For high values of $\kappa$, we see that the color hypothesis $H_\text{color}$ does not model the groups well compared to the hypotheses $H_\text{link}$ and $H_\text{link-colord}$ that assume equal likelihood of all links.

Overall, \approachname{} yields results that are in line with the actual generation process of the datasets.
Our approach thus allows to derive information about the quality of the group assignments as well as the transition behavior within the groups.
The strongly diverging characteristics of the different hypotheses illustrates the flexibility of \approachname{}.
\vspace{-.5em}
\subsection{Synthetic Datasets: Probabilistic Group Assignments}
\label{sec:synthprob}
\vspace{-.5em}
So far, we have only considered \emph{deterministic} group assignment probabilities in the experiments, i.e., assigning transitions to a single group by only using binary probabilities: $\gamma_{g|t} \in \{0,1\}$.
However, there is a wide variety of situations where it is useful to consider probabilistic group assignments or fuzzy walkers, e.g., when considering smooth behavior transitions between different times of a day, when transitions are assigned to groups by an uncertain classifier, or when walkers randomly choose between different movement patterns.
Here, we explore probabilistic group assignments in a synthetic dataset.
For a real world example of an uncertain classifier, see \Cref{sec:flickr}.
\fle{Also in Wikispeedia?}
\begin{figure}[t!]
    \centering
        \includegraphics[scale=0.5]{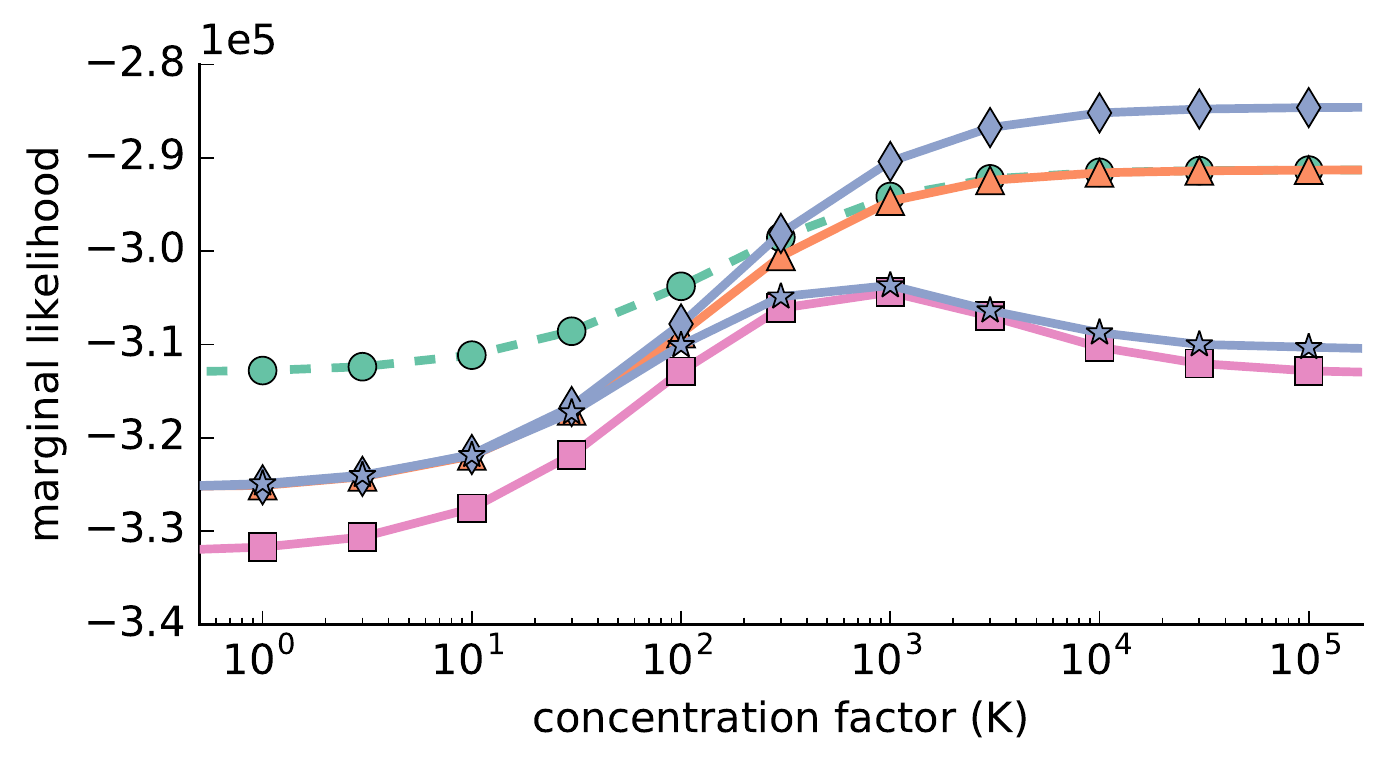}
        \includegraphics[scale=0.5]{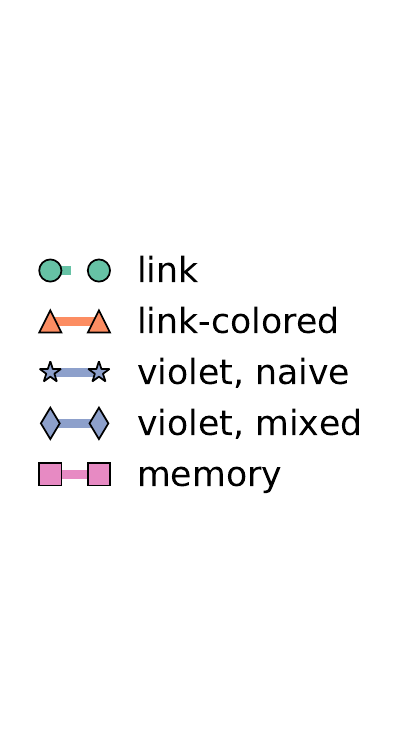}
    \caption{\textbf{Probabilistic group assignments on synthetic data.} The \emph{violet, mixed} hypothesis, using probabilistic group assignment probabilities, is the most plausible one for increasing concentration factors as it directly models the processes underlying the data. The \emph{violet, naive} hypothesis illustrates the integral role of the mixing step, as skipping it significantly reduces the performance of a hypothesis even though the underlying processes were correctly understood. Further details are discussed in \Cref{sec:synthprob_results}.
    }
    \vspace{-1em}
    \label{fig:synthetic-violet}
\end{figure}

\para{Dataset. }
\label{sec:synthprob_data}
We use the same underlying network as in the previous example to construct a dataset.
However, instead of ``red'' and ``blue'' walkers, the sequences are now generated by walkers with ``mixed colors'', called \emph{violet walkers},
i.e., the walkers randomly choose to walk according to the red $\mtheta_\text{red}$ or to the blue $\mtheta_\text{blue}$ transition probability matrix at each step.
For example, a violet walker $w$ associated with a shade of violet $s_w = 0.3$ will choose to be a red walker for 30\%, and a blue walker for 70\% of her transitions.
We create a dataset $D_\text{violet}$ of 10,000 walkers that each perform 10 transitions.
We assign a shade of violet $s_w$ to each walker $w$, which we draw from a Beta distribution $s_w \thicksim Beta(1,1)$.
Before each transition of a walker, she randomly draws a color $c \in \{\text{red}, \text{blue}\}$ according to her shade of violet $s_w$ using a Bernoulli distribution $c \thicksim Bernoulli(s_w)$.
Then, she uses the respective transition matrix $\mtheta_\text{red}$ or $\mtheta_\text{blue}$ dependent on the chosen color $c$ to determine her next destination.

\para{Hypotheses. }
\label{sec:synthprob_hypo}
As hypotheses, we define $H_\text{link}$, $H_\text{link-colored}$ and $H_\text{memory}$ analogously to \Cref{sec:synth_hypo}.
In addition, we introduce a hypothesis $H_\text{violet} = (\mgamma_\text{violet}, \mphi_\text{violet})$ specifically tailored to violet walkers. 
Thus, we define the group dependent transition probabilities as $\mphi_\text{violet} = (\mtheta_\text{red}, \mtheta_\text{blue})$.
Now, violet walkers choose transition probability matrices \emph{probabilistically} dependent on their shade of violet.
Using our MTMC scheme, this can be modeled by setting the corresponding group assignment probabilities according to a walker's shade of violet $s_w$: $\gamma_{g|t_w} = (s_w, 1 - s_w)$, where $t_w$ represents a transition from a specific walker $w$. 

\para{Results. }
\label{sec:synthprob_results}
The results are shown in \Cref{fig:synthetic-violet}.
The first observation is that the violet hypothesis $H_\text{violet}$ (mixed) works best for increasing concentration factors.
Note that we consider two variants of the violet hypothesis, one (\emph{violet, mixed}) elicited using the mixing method proposed in \Cref{sec:eliciting} and one (\emph{violet, naive}) elicited as if it was a deterministic hypothesis.
The results show that the mixing step is an integral part of \approachname{}, as skipping it significantly reduces the performance of the heterogeneous hypothesis even though the underlying processes were correctly understood. 

As for the other hypotheses from \Cref{fig:synthetic-violet},
the \emph{link} hypothesis works best.
This is because, generally, a perfectly violet walker ($s_w = 0.5$) behaves exactly like a link walker.
This also explains the differing results for lower concentration factors:
The grouping introduced by the violet hypothesis injects complexity which is not splitting transitions in a manner that can easily be explained.
Thus, for low concentration factors, which imply a large uncertainty in the hypothesis, this reduces the plausibility of the more complex hypothesis. 
However, with growing concentration factors the better modeling of the transition probabilities justifies the added complexity making the violet (\emph{mixed}) hypothesis the most plausible one.

With regard to the increased complexity, the colored (heterogeneous) link hypothesis (\emph{link-colored}) has the same disadvantage as the violet hypothesis; consequently, it is inferior to the homogeneous link hypothesis.
The memory hypothesis has the lowest plausibility as it does not reflect the generative process of the dataset and introduces three groups instead of just two.

Overall this example shows that, by using \approachname{}, heterogeneous data can be modeled accurately and that the mixing procedure for eliciting probabilistic hypotheses as introduced in \Cref{sec:eliciting} is an integral part of the approach.

\subsection{The Wikispeedia dataset}
\label{sec:wiki}

Wikispeedia~\cite{west2012human} is a game in which players aim to find a short path from a randomly given start article to a randomly given target article within Wikipedia by only navigating the available hyperlinks. 
In the context of this game, the authors have hypothesized that ``humans navigate more strongly according to degree in the early game phase, when finding a good hub is important [in order to be able to increase the amount of reachable concepts], and more strongly according to textual similarity later on, in the homing-in phase [when trying to find the actual target concept]''.
Here, we confirm this hypothesis using \approachname{}.

\para{Data.}
Wikispeedia is based on a subset of 4,600 Wikipedia articles (from the 4,600-article CD version of ``Wikipedia for Schools''\footnote{available at \url{schools-wikipedia.org} (version of 2007)}).
A corresponding dataset~\cite{west2009wikispeedia} is freely available\furl{https://snap.stanford.edu/data/wikispeedia.html}.
It consists of the plain text of each article, the link network, and a set of click sequences (including back clicks) created by humans playing the game.
Like West et al.~\cite{west2012human}, we remove back clicks (but keep the corresponding forward clicks which are undone by these back clicks) and then only keep click sequences of length 3 to 8 (number of clicks). 
The resulting dataset consists of over 25,000 click sequences with a mean length of 5.6. 

\para{Hypotheses. }
To investigate the hypothesis from~\cite{west2012human}, we consider two transition probability matrices:
$\mphi_\text{deg}$ represents the hypothesis that people are trying to get to hubs in order to increase the number of concepts they can reach.
Thus, if a link between a source article to a destination article exists, we set the belief in the corresponding transition proportional to the degree of the destination state (calculated as the sum of its in- and out-going links); and zero otherwise.
Second, the transition probability matrix $\mphi_\text{sim}$ assumes a higher transition probability if there is a strong textual similarity between two articles.
Again, we set the transition probability to $0$ if there is no link between two articles.
Otherwise, we set the belief in a transition proportional to the cosine similarity $cos(i,j)$ with respect to the corresponding \textit{tf-idf} vectors. For that, we removed words with a document frequency of over 80\%  and applied sublinear scaling to the tf values. 
\footnote{Differing from our approach, West et al. \cite{west2009wikispeedia} use the similarity between the clicked article and the target concept $cos(i,t)$, but report that along the game progress, the similarity of the current and the clicked/next article is qualitatively similar. Thus, we use the latter approach since we can only use information from already visited states, not future states.}
For comparison, we additionally consider 
the link matrix $\mphi_\text{link}$ that expresses equal belief in all transitions for which a link exists. 

Now, the first three hypotheses are homogeneous hypotheses assigning transitions to a single group similar to \Cref{fig:hypotheses}:  $H_\text{link} = (\mgamma_\text{one}, \mphi_{link})$, $H_\text{deg} = (\mgamma_\text{one}, \mphi_{deg})$, $H_\text{sim} = (\mgamma_\text{one}, \mphi_{sim})$.
Furthermore, $H_\text{deg,sim}$ and $H_\text{sim,deg}$ are heterogeneous hypotheses that group transitions based on their position on the trail of articles left by users playing the game.
In particular, the first two transitions are assigned to the ``initial phase'', and the rest of the transitions are assigned to the ``homing-in phase''.
We name the corresponding group assignment probabilities $\mgamma_\text{phases}$.
The heterogeneous hypotheses are then defined as: $H_\text{deg,sim} = (\mgamma_\text{phases}, (\mphi_{deg}, \mphi_{sim}))$ and $H_\text{deg,sim} = (\mgamma_\text{phases}, (\mphi_{sim}, \mphi_{deg}))$ assuming the degree and the similarity transition probability matrices to explain the ``initial phase'', respectively. 

\para{Results. }
\begin{figure}[t!]
    \centering
    \subfigure {
        \includegraphics[scale=0.50]{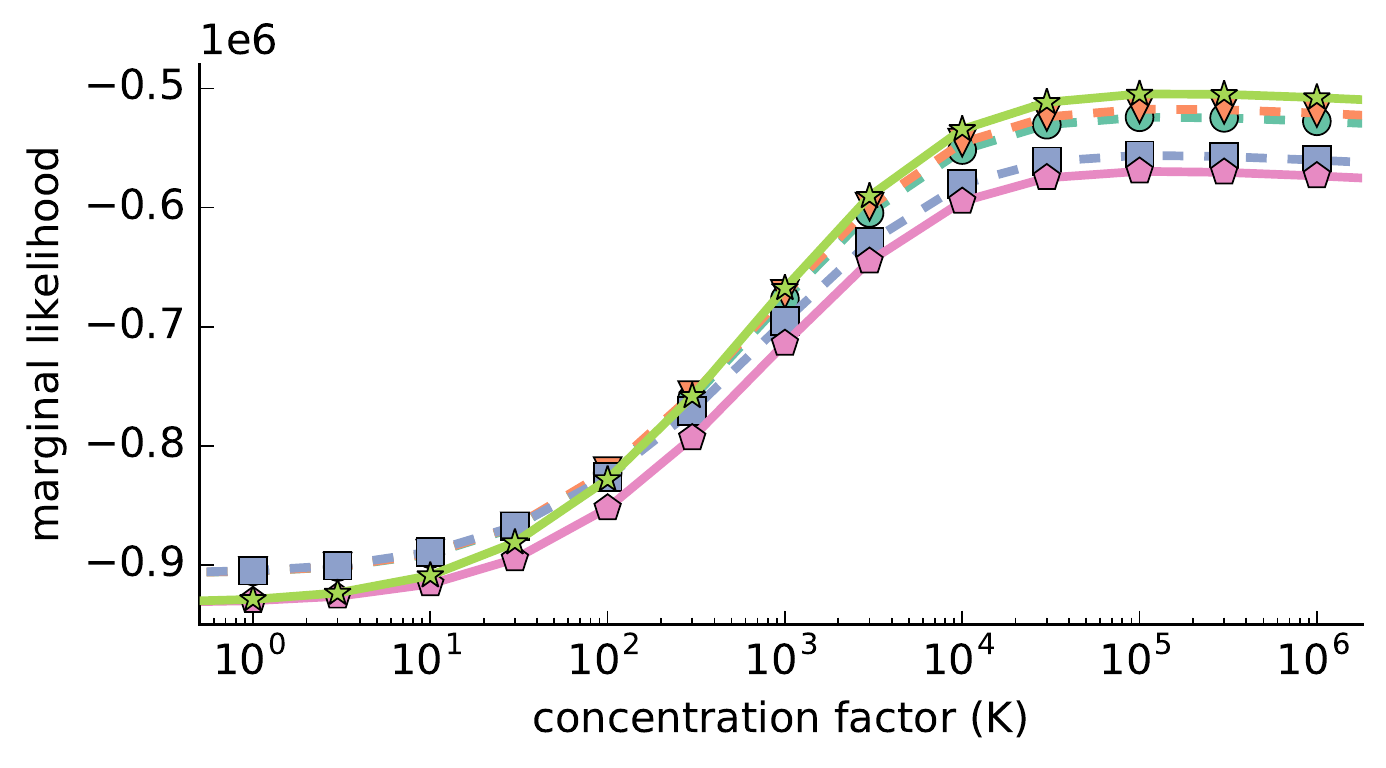}
    }
    \subfigure {
        \includegraphics[scale=0.50]{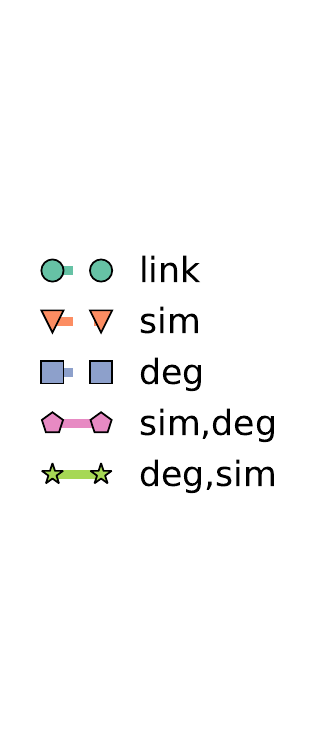}
    }
    \caption{\textbf{Wikispeedia results.}
    For the game Wikispeedia, players try to quickly navigate from one article to another using the underlying link structure of Wikipedia. 
    One hypothesis (\emph{deg,sim}) is that players will first navigate to articles with a large degree, and then ``home-in'' on their target using similarity based navigation.
    The graph shows the results of modeling this heterogeneous hypothesis using the MTMC model by splitting each click sequence after their second click.
    We also compare against several other homogeneous as well as heterogeneous hypotheses.
    Overall, of all the considered hypotheses, the heterogeneous \emph{deg,sim}-hypothesis works best (for growing concentration factors), even though the initial split (at concentration factor $\kappa = 0$) is not inherently advantageous. 
    For details, see \Cref{sec:wiki}.
    Note that, while the differences visually appear to be marginal in the plot, they are decisive (cf. \Cref{sec:example}).}
    \vspace{-2em}
    \label{fig:wiki}
\end{figure}
\Cref{fig:wiki} shows that, as literature hypothesized, the heterogeneous hypothesis $H_\text{deg,sim}$ explains the  navigational behavior of players  better than all other considered hypotheses.
While the additional variables introduced by the split (by means of Occam's Razor) result in lower marginal likelihoods compared to the homogeneous hypotheses for weak believes (low values of the parameter $\kappa$), it becomes apparent that the transition probability matrices of $H_\text{deg,sim}$ are modeling the corresponding movement behavior in each group better than the single transition probability matrix of the homogeneous hypotheses.
At the same time, the ``opposite'' hypothesis $H_\text{sim,deg}$ results in the lowest ML values, even though it uses the same split as $H_\text{deg,sim}$.
Among the homogeneous hypotheses, the similarity based hypothesis is the most plausible.
By contrast, as it yields rather low ML values, the degree hypothesis $H_\text{deg}$ seems to be a very specialized hypothesis, which is applicable only for a specific subset of transitions; such as the first transitions in each sequence.

Overall, this example demonstrates the applicability of \approachname{} to a real world scenario. We also see that a more fine-grained hypothesis may explain observed sequential data better than using a single, overly general hypothesis.

% \mbe{blah blah}
\subsection{The Flickr dataset}
\label{sec:flickr}

\mbe{Initial text ... needs rework. Please check consistency 2.5/5.0 VS 5.0/2.5}
Finally, we investigate geo-spatial trails obtained from the photo-sharing platform Flickr\furl{https://www.flickr.com/}.

\para{Dataset. }
\label{sec:flickr_data}
As data in this setting we employ a dataset from previous work~\cite{lemmerich2016subtrails}.
It contains all Flickr photos from the years 2010 to 2014 with geo-spatial information (i.e., latitude and longitude) at street-level accuracy in Manhattan. 
We mapped each photo according to its geo-location to one of the 288 census tracts (administrative units) that we use as state space in our model (see also \cite{gambs2010showme}). 
Then, for each user, we built a sequence of different tracts she has taken photos at (excluding self-transitions). 
Thus, we know the start and end date for each user sequence.
The final dataset contains 386,981 transitions overall.

\para{Hypotheses. }
\label{sec:flickr_data_hypo}
In previous research \cite{becker2015photowalking}, we found that a combination of spatial proximity to the current state and to points of interest (PoIs) is the best explanation for the transitions of Flickr users.
However, in some settings, proximity to the current state is more relevant, while in others, larger, spatial variances lead to better results.
Accordingly, we built two different transition probability matrices that we call $\mphi_\text{near}$ and $\mphi_\text{far}$, which feature different parametrizations of the proximity/POI hypothesis.
In particular, we set the influence radius of PoIs to $400 m$ and the standard deviation of the proximity factor to $2.5 km$ ($\mphi_\text{near}$) and $5.0 km$ ($\mphi_\text{far}$). 
For details, we refer to \cite{becker2015photowalking}.

In this paper, we aim to extend the previous study by taking into account whether a user is a \emph{tourist} or a \emph{local}.
To classify users as tourists or locals, we use the time difference between their first and their last photo in the data, cf. \cite{dechoudhury2010automatic}.
In that regard, we consider different group assignments:
a) a baseline $\mgamma_\text{one}$ that puts all transitions into one group,
b) deterministic grouping $\mgamma_\text{det}$ by defining tourists as users with a trail duration of 21 or less days, and 
c) a smooth distinction between tourists and locals around 21 days by using a sigmoid function $sig(t) = \nicefrac{1}{1 + e^{-t}}$ resulting in probabilistic group assignments $\mgamma_\text{prob}$.

We combine these three group assignments and transition probability matrices to form five hypotheses: 
(i) $H_\text{near}= 
    (\mgamma_\text{one}, \mphi_\text{near})$, 
(ii) $H_\text{far}= 
    (\mgamma_\text{one}, \mphi_\text{far})$, 
(iii)  $H_\text{det: tourist=near} = 
    (\mgamma_\text{det}, (\mphi_\text{near}, \mphi_\text{far}))$, 
(iv) $H_\text{prob: tourist=near}= 
    (\mgamma_\text{prob}, (\mphi_\text{near}, \mphi_\text{far}))$, and 
(v) $H_\text{prob: tourist=far}= 
    (\mgamma_\text{prob}, (\mphi_\text{far}, \mphi_\text{near}))$.
For example, the last hypothesis $H_\text{prob: tourist=far}$ expresses a belief that there are two groups --- locals and tourists --- in the data, and the longer the sequence of a user is (in days), the  more likely she is to be a local. Furthermore, this hypothesis assumes that tourist are more likely to have a longer distance to the next photo location than locals.
We additionally added a homogeneous uniform hypothesis as a baseline that assumes that all transitions are equally likely and that no groups exist.

\para{Results. }
\label{sec:flickr_data_results}
\begin{figure}[t!]
    \centering
    \subfigure {
        \label{fig:2_all}
        \includegraphics[scale=0.5]{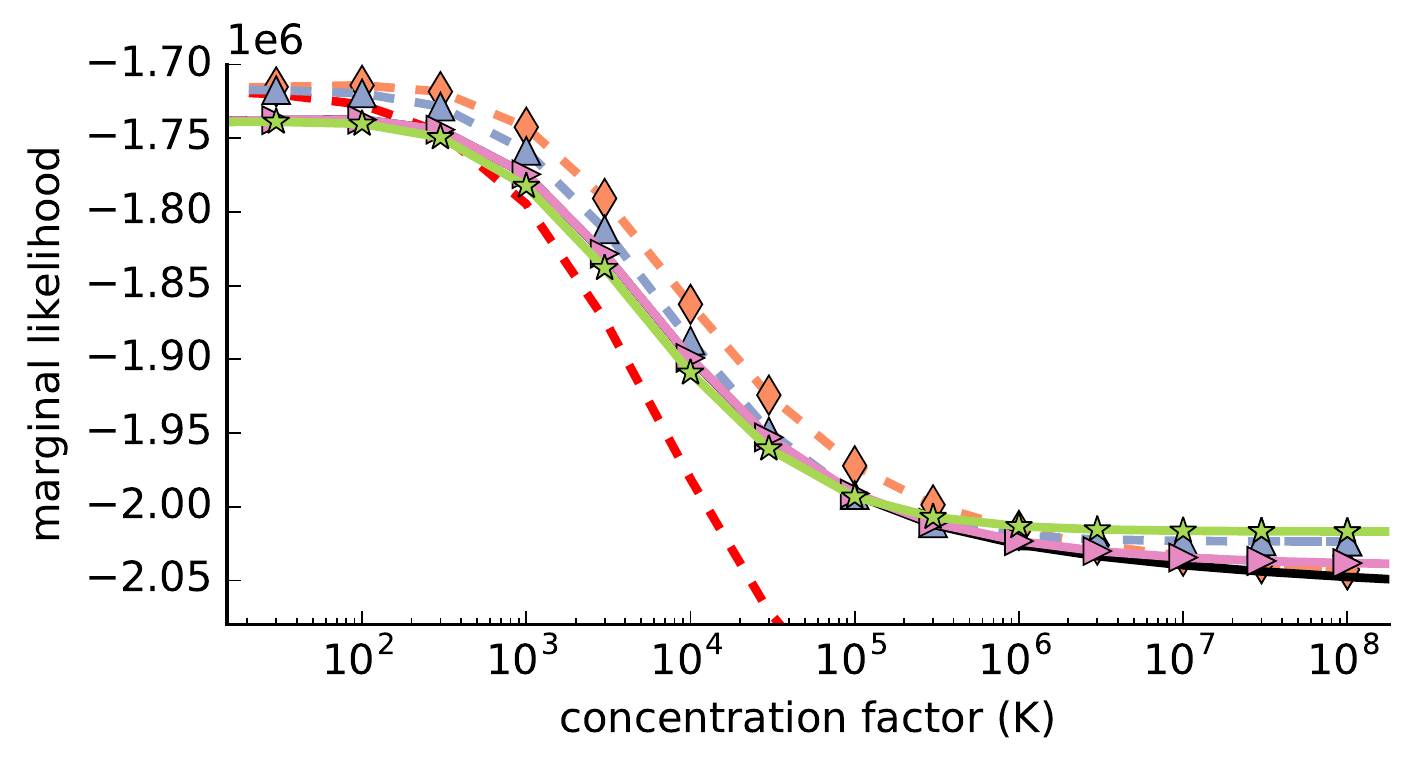}
    }
    \subfigure {
        \label{fig:rnd_all}
        \includegraphics[scale=0.5]{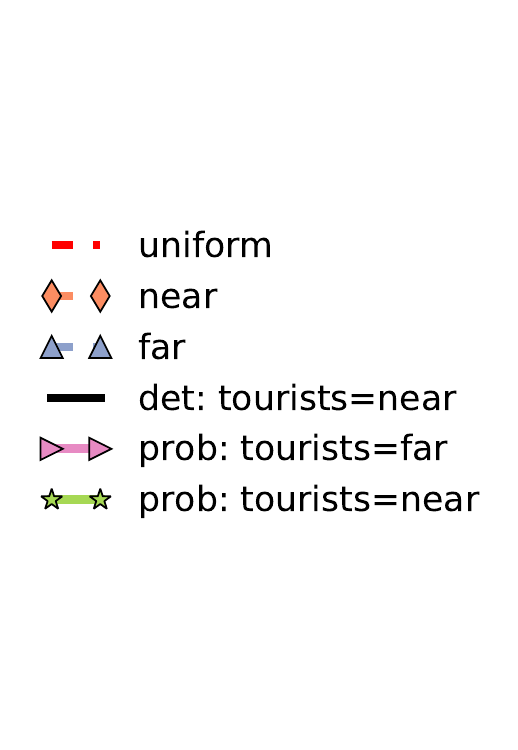}
    }
    \vspace{-2mm}
    \caption{\textbf{Flickr results.} 
    We model the navigation behavior between tracts in Manhattan based on photo trails on the social photo-sharing platform Flickr.
    Overall, we have not found hypotheses explaining the data well as indicated by the strongly decreasing marginal likelihoods.
    However, those we evaluated are better than the baseline, i.e., the uniform hypothesis. The best one (\emph{prob: tourists=near}) assumes that tourists are more prone to move to \emph{close by} tracts than locals.
    Here, MTMC allows for modelling uncertain classification of tourists which covers the underlying processes better than a deterministic group assignment (\emph{det: tourists=near}). 
    }
    \vspace{-1em}
    \label{fig:flickr}
\end{figure}
\Cref{fig:flickr} shows the results.
Obviously, the uniform hypothesis is substantially less plausible than all proximity/PoI-based hypotheses.
Among the latter, we see that for smaller concentration factors homogeneous groupings perform better, which indicates that in general the split into tourists and locals by itself does not produce particularly distinct movement behavior.
However, for increasing concentration factors $\kappa$, it turns out that the hypothesis $H_\text{prob: tourist=near}$ works best, i.e., by using probabilistic group assignments and expressesing the belief that tourists take their next photo at a more near-by location with a close PoI while locals choose locations with higher distance more often.
By contrast, a deterministic split $\mgamma_\text{det}$ does not cover the uncertainty of classifying tourists and locals.

Overall, this example illustrates how \approachname with probabilistic group assignments enables more fine-grained analyses of sequential data.%, and how the use of probabilistic group assignments (i.e., uncertain classification of tourists and locals using the sigmoid function) can be applied in order to capture a real world scenario.  
\vspace{-3mm}
\section{Discussion}
\label{sec:discussion}
\vspace{-3mm}
    
With \approachname, we have proposed a powerful approach to formulate and compare hypotheses about heterogeneous sequence data.
In this section, we discuss some alternative choices as well as possible misunderstandings and shortcomings of our method.

\para{Top-down vs. bottom-up.}
\begin{revised}
\approachname{} is a \emph{top-down} approach ---
also called a \emph{deductive} approach in certain contexts
\cite{trochim2001research,herr2008sourcebook,blackstone2012sociological}
--- meaning that it takes a set of hypotheses based on  \emph{ideas} and
\emph{theories} from the application domain as input and compares them using some observed data.
While the corresponding results also give an indication of the predictive potential 
of hypotheses, we do not fit them to the data.
For utilizing the data to learn models that excel at prediction, a multitude
of other, more specialized methods are available, 
e.g., ~\cite{yang2014finding,figueiredo2016mining,laxman2008stream}.
Note, that these methods usually do not yield directly interpretable results.
If they do (e.g., \cite{figueiredo2016mining}), they can be used in a
\emph{bottom-up} setting --- sometimes also called an
\emph{inductive} \cite{trochim2001research,blackstone2012sociological} setting ---, which takes the opposite approach than \approachname{}:
bottom-up methods use observations to extract patterns or regularities 
from which \emph{new} hypotheses or theories can be derived.
The same is true for other specialized approaches, e.g., for segmentation, 
labeling, or clustering~\cite{ponte1997text,blei2001topic,wallach2006topic,van1998text,fox2010bayesian}: 
while they can be facilitated or extended to \emph{uncover new} interesting
heterogeneous patterns by examining their latent structures, applying them results 
in a bottom-up approach which \emph{yields new} hypotheses instead of comparing existing ones.
\end{revised}

\begin{revised}
\para{Extensions and alternative approaches.}
While \approachname{} provides a very flexible and easy to understand framework for specifying and comparing hypotheses, there is a variety of possible extensions and alternative approaches.
For example, in this work, we employ priors for transition probabilities, but specify group assignment probabilities directly and fixed, which somewhat forces the user to be very specific with regard to group assignments.
In contrast, using a flat prior over group assignments, the user could compare hypotheses that introduce groups of transition probabilities without having to specify which transition belongs to which process.
Also, \approachname{} can not directly express dependencies between the groups of the transitions within a sequence (e.g., stickiness \cite{fox2010bayesian,wetzels2016bayesian}), as for example possible in Markov switching processes such as the Hidden Markov model. 
That is, while we can construct hypotheses in a way such that group assignment probabilities are derived by Hidden Markov structures, hidden state dependencies can not be explicitly modelled.
We could resolve this by using more complex models for sequential data.
This, however, would come at the cost of substantially increased efforts for specifying model parameters in the hypotheses, especially considering the wide range of incorporated background knowledge.
Overall, \approachname{} tries to balance the amount of parameters required to formulate a hypothesis against expressiveness.
Nevertheless, we acknowledge the potential of formulating more complex dependencies with the help of more complex models, especially when considering the possibility of flat/uninformed priors over certain parameter groups, but leave further studies to future work.
\end{revised}

\para{\approachname{} vs. separate HypTrails comparisons.}
A simplistic alternative to our approach could be to apply the original HypTrails method for homogeneous data separately to the groups of a hypothesis.
This, however, is limited to deterministic group assignments and does not allow to compare hypotheses with different group assignments (or no group assignments at all).
In addition, \approachname{} provides a theoretical background on how to aggregate results for the individual groups, i.e., by multiplying their marginal likelihood.

\para{Using different strengths of belief.}
We are using different strengths of belief (i.e., concentration factors $\kappa$) in order to study different properties of our hypotheses. 
Calculating the marginal likelihood for very large concentration factors $\kappa$ approximates the likelihood of the model for fixed parameters, which is commonly used to compare parameter settings in frequentist statistics (e.g., via likelihood ratio test).
However, by also investigating lower concentration factors, we obtain additional information on the quality of the group assignments (cf. \Cref{sec:example}).
Furthermore, our approach enables the observation of the dynamics for growing concentration factors, which allows us to
judge whether a hypothesis covers predominant factors of the underlying processes generating the sequential data.
Thus, we believe that the analysis based on different concentration factors can yield a more detailed comparison of hypotheses than other, one dimensional measures, such as the model likelihood, which is included in our approach as a special case and shown on the right-hand side of our result plots.

\begin{revised}
Nevertheless, we acknowledge that it may be useful to derive a single number by which hypotheses can be compared.
To achieve this we could either set a fixed $\kappa$ according to some background information or, in a more Bayesian way, we could treat the concentration parameter $\kappa$ as a free parameter and marginalize over it.
This, however, would require specifying a prior over this free parameter, which is inherently a difficult choice.
As a simple solution, we propose to compute the average marginal likelihood over a set of $\kappa$ values. This is equivalent to a prior that regards these values as equally likely.
Overall, summarizing result curves into a single value in this way requires additional task-dependent choices and comes with a loss of information in the result on the one hand, but allows for a more compact representation of results on the other hand.
Developing guidelines for choosing appropriate priors over $\kappa$ remains an open issue for future work.
\end{revised}

\para{Efficiency and convergence.}
In the general case, the marginal likelihood of the MTMC model has to be approximated.
While the method from \Cref{sec:inference} has converged quickly ($\ll 50$ iterations) so that we were able to calculate our results on regular consumer hardware in a few hours, parallelizations along the lines of \cite{becker2016sparktrails} may be useful for larger datasets.
We have also experimented with other methods for approximating the marginal likelihood such as \cite{chib1995marginal}, but have found irregularities in the convergence behavior.
Further studies may address both, the parallelization of our method and exploring other approximation schemes.

\begin{revised}
\para{Multiple comparisons. }
Our approach enables the comparison of multiple hypotheses against each other.
In that direction, it can also be checked whether one of the hypotheses performs better than a simple baseline hypothesis such as the uniform hypothesis.
If many hypotheses are tested in this way, then the multiple comparison problem should be taken into account. 
That is, even if hypotheses are generated purely random, some of them would appear to be statistically significantly better than the baseline, cf.~\cite{benjamini1995controlling}.
Although our approach is in principle affected by this problem, we see this issue as non-crucial in our setting as (i) the main goal of our approach is not to show whether one of our hypotheses can beat a baseline, but to compare hypotheses against each other (pairwise) and (ii) we use only a comparatively small set of hand-elicited hypotheses in our comparisons.
Apart from that, there is intense discussion how multiple comparisons are to be viewed from a Bayesian perspective, see for example~\cite{goodman1998multiple,gelman2012we}.
Nonetheless, exploring the challenges of multiple comparisons will be an issue that we will study more in-depth in future work.
\end{revised}

\section{Related work}
\label{sec:relatedwork}

In this section, we provide an overview of related work on Markov chain models, their applications (focusing on the Web context), and respective extensions. Further discussions and elaborations of related work have been captured throughout the course of this work.

(First-order) Markov chain models have been first introduced by A. A. Markov in 1913 
\cite{markov2006example}. 
Several adaptations of the Markov chain model have been proposed, such as the so-called higher-order Markov chain models \cite{singer2014detecting}, Hidden Markov models \cite{rabiner1986introduction}, or mixtures of Markov models \cite{poulsen1990mixed,smith1997markov}.
Historically, Markov chain models have been applied in many diverse settings due to their simplicity and generality. Examples include textual data modeling \cite{markov2006example}, weather data modeling \cite{gabriel1962markov}, C.E. Shannon's take on information theory \cite{shannon2001mathematical}, or the application of modeling Web navigation leading to the PageRank algorithm \cite{page1999pagerank}. 
A historical summary of Markov chain models and their applications can be found in \cite{hayes2013first}.  

In this work, we have focused on applications in the Web context.\footnote{Note that our approach can also be applied to very different settings in a straight-forward manner.} 
This line of research has been tackled in a multitude of studies. 
For example, early work by Catledge and Pitkow \cite{catledge} investigated human navigation on WWW pages.
Subsequent studies have further demonstrated that Web navigation is guided by certain regularities \cite{huberman,pirolli99,chi2001,west2012human,walk2014cikm}.
Prominent theories are, for example, that humans prefer to transition between semantically similar concepts \cite{chalmers1998order,brumby,west2012human}, or the so-called \emph{information foraging theory} \cite{pirolli99,chi2001} postulating that human behavior in an information environment on the Web 
is guided by \emph{information scent}. 
Among many others, further studies have focused on sequence prediction \cite{laxman2008stream,asahara2011pedestrian,noulas2012mining,figueiredo2016tribeflow}, the recommendation of travel routes \cite{dechoudhury2010automatic}, search trails \cite{white2010assessing}, or the study of music sequences~\cite{baccigalupo2006case,figueiredo2016mining}.

Motivated by this large array of hypotheses about sequential behavior, HypTrails~\cite{singer2015} was proposed for comparing the plausibility of hypotheses on sequential data. \approachname, as introduced in this paper, builds on HypTrails and addresses one of its main issues, namely allowing to model and compare hypotheses about \emph{heterogeneous} sequence data.

\begin{revised}
The model we employ in this paper is related to previously proposed extensions of Markov chains.
One prominent example are mixture models~\cite{smith1997markov}.
In that direction, the \emph{Mixed Markov chain} model has been studied by Poulsen \cite{poulsen1990mixed} in the context of customer behavior segmentation.
Poulsen, however, defines group memberships on a sequence level, not on a transition level sacrificing some of the expressiveness incorporated into \approachname{}.
Similar group memberships are used by Rendle et al. who factorize Markov chains~\cite{rendle2010factorizing} and by Gupta et al. who reconstruct mixtures of Markov chains \cite{gupta2016mixtures}.
To our knowledge, these models have not been employed for the comparison of hypotheses so far.
Additionally, the expressiveness of these models is limited, i.e., not all group assignments of the hypotheses featured in this paper could be expressed with these models.
Another set of Markov chain extensions related to our approach is the class of Markov switching processes~\cite{rabiner1986introduction,fox2010bayesian} which model observations dependent on hidden Markovian dependency structures.
Some classic instances in this class are the Hidden Markov Model (HMM)~\cite{rabiner1986introduction}, the Factorial HMM \cite{ghahramani1997factorial} or the the Auto-Regressive HMM \cite{hamilton1990analysis} (also see \cite{murphy2002dynamic} for further extensions). 
There are also methods based on, or related to, these methods which are used for prediction, clustering or segmentation  \cite{figueiredo2016tribeflow,fruhwirth2008model,matsubara2014autoplait,goldwater2007fully}, including, e.g., Bayesian nonparametric methods \cite{teh2006hierarchical,fox2010bayesian} which adjust their complexity based on the data.
However, such methods fit models to the data, i.e., they learn model parameters.
Sometimes these model parameters can be used to \emph{find new} hypotheses (as opposed to comparing \emph{existing} ones).
While, e.g, Hidden Markov models have been applied to compare streaky behavior with a baseline model \cite{wetzels2016bayesian}, to best of the authors knowledge, there are no general approaches to apply Markov switching processes in a top-down manner in the context of background data.
For a more detailed discussion on the difference between top-down and bottom-up approaches please see \Cref{sec:discussion}.
\end{revised}

Statistical methods for comparing the fits of different Markov chain models have been summarized in \cite{singer2014detecting} and include likelihood ratio tests, information-theoretic AIC, BIC or DIC approaches, or the Bayes factor. These methods have been utilized, e.g., for comparing the fit of nested, higher-order Markov chain models that relax the basic assumption of the Markovian property and allow for longer memory fits. In this work, we focus on comparing fits by using marginal likelihoods and Bayes factors \cite{strelioff2007inferring}; these have the advantage of an automatic built-in Occam's razor balancing the goodness of fit with complexity \cite{kass1995bayes}. Additionally, instead of only using a flat Dirichlet prior, we also utilize the sensitivity of the marginal likelihood on the prior for comparing theory-induced hypotheses within the Bayesian framework---as advocated, e.g., in \cite{rouder2009bayesian,vanpaemel2010prior,kruschke2014doing}---following the HypTrails approach elaborated in \cite{singer2015}.
To the authors' knowledge, there exist no previous approaches for the comparison of hypotheses about transition behavior that differentiate between several groups contained in the data.
This is in line with a general trend towards Bayesian methods for data analysis~\cite{kruschke2013bayesian,benavoli2014bayesian}.

\vspace{-2mm}
\section{Conclusions}
\label{sec:conclusion}
\vspace{-2mm}
In this paper, we have introduced \approachname{}, a Bayesian method for comparing hypotheses about the underlying processes of heterogeneous sequence data.
\approachname{} incorporates (i) a method for formulating heterogeneous hypotheses using (ii) the \emph{Mixed Transition Markov Chain} (MTMC) model, which enables specifying individual hypotheses for very flexible subsets of transitions, i.e., with regard to certain user groups, state properties, or the set of antecedent transitions.
Furthermore, (iii) we introduced methods for eliciting hypotheses as parameters for this model, (iv) showed how to calculate the marginal likelihood, and (v) provided some guidance on how result plots can be interpreted to compare the corresponding hypotheses.
The benefits of our approach were demonstrated on synthetic datasets, and we gave application examples with real-world data.
Overall, this work enables a novel kind of analysis for studying sequence data in many application areas.

In the future, we may explore our method in additional real-world applications, such as investigating the movement of (groups of) Flickr users, cf.~\cite{becker2015photowalking}, or  studying groups of editors in Wikipedia.
\begin{revised}
Furthermore, more complex priors or hierarchical models may allow for more powerful ways of expressing hypotheses.
\end{revised}

\section{Acknowledgements}
This work was partially funded by the BMBF project Kallimachos and the DFG German Science Fund research projects PoSTs II and p2map.

\small
\sloppy
\bibliographystyle{plain}
\bibliography{bib}

\appendix
\clearpage
\section{Derivation of the marginal likelihood of MTMC}
\label{app:ml}
Given the generative process from \Cref{sec:model} and by exploiting the fact that the transition probabilities $\mtheta_g$ for each group $g$ as well as the group assignment probabilities $\gamma_{g|t_k}$ for each transition $t_k$ are independent, we can write the marginal likelihood of MTMC as follows:
\begin{align*}
\Pr(D|H)
& = \int\underbrace{\Pr(D | \mtheta,\mgamma)}_{\text{likelihood}} ~ \underbrace{\Pr(\mtheta|\malpha)}_{\text{prior}} ~ d\mtheta \\
& = \numberthis \label{eq:ml-initial} \int \underbrace{\prod_{t_k \in D} \sum_{g \in G} \gamma_{g|t_k}  \theta_{i_k,j_k|g}}_{\Pr(D | \mtheta,\mgamma)}  \underbrace{\prod_{g \in G} \Pr(\mtheta_g|\malpha_g)}_{\Pr(\mtheta|\malpha)}  \prod_{g \in G} d\mtheta_g
\end{align*}
To solve this integral we take a similar path as in the homogeneous case (cf. \cite{singer2015}).
Thus, we need to get the grouping out of the integral.
First, we focus on the likelihood $\Pr(D | \mtheta,\mgamma)$ where we extend the multiplication over all transitions resulting in an outer sum over all possible group assignments:
\begin{align*}
\Pr(D|\mtheta, \mgamma) & = \prod_{t_k \in D} \sum_{g \in G} \gamma_{g|t} \theta_{i_k,j_k|g}\\
& = \sum_{\substack{\omega \in \Omega \\ \Omega = \{\{(t_1, g_1), ..., (t_m, g_m)\} | (g_1, ..., g_m) \in G^{|D|}\}}} \prod_{(t_k, g_k) \in \omega} \gamma_{g_k|t_k}  \theta_{i_k,j_k|g_k} \\
& = \sum_{\omega \in \Omega} \underbrace{\prod_{(t_k, g_k) \in \omega} \gamma_{g_k|t_k}}_{p_\omega}  \prod_{(t_k, g_k) \in \omega}  \theta_{i_k,j_k|g_k} \\
& = \numberthis \label{eq:likelihood-reformulated} \sum_{\omega \in \Omega} p_\omega  \prod_{g \in G} \prod_{s_i,s_j \in S}  \theta_{i,j|g}^{n_{i,j|g,\omega}} 
\end{align*}
Here, each $\omega$ represents a single, fixed group assignment of the set of transitions in $D$ where 
the set of all possible group assignments $\omega$ is defined as $\Omega = \{\{(t_1, g_1), ..., (t_m, g_m)\} | (g_1, ..., g_m) \in G^{|D|}\}$.
Furthermore, $p_\omega$ represents the probability of the respective group assignment $\omega \in \Omega$.
Finally, $n_{i,j|g,\omega}$ denotes the number of transitions from state $s_i$ to state $s_j$ given the group $g$ and the group assignment $\omega$.
What we observe is that, given a specific group assignment $\omega$, the likelihood is the same as the likelihood in \cite{singer2015}. 

We now substitute the likelihood $\Pr(D|\mtheta, \mgamma)$ in \Cref{eq:ml-initial} with this reformulated likelihood (\Cref{eq:likelihood-reformulated}) and write the priors for the group dependent transition probabilities $\Pr(\mtheta_g|\malpha_g)$ based on the multivariate beta function.
Then, we can calculate the marginal likelihood $\Pr(D|H)$ by taking advantage of the independence of the transition probabilities $\mtheta_g$ between groups $g \in G$ and source states $s \in S$ as well as the independence of group assignment probabilities $\gamma_{g_k|t_k}$ between transitions $t_k \in D$:
\begin{align*}
\Pr(D|H)
&= \int \underbrace{\sum_{\omega \in \Omega} p_\omega  \prod_{g \in G} \prod_{s_i,s_j \in S}  \theta_{i,j|g}^{n_{i,j|g,\omega}} }_{\Pr(D|\mtheta, \mgamma)}  \prod_{g \in G}  \underbrace{\prod_{s_i \in S} \frac{1}{B(\valpha_{s_i|g})} \prod_{s_j \in S} \theta_{i,j|g}^{\alpha_{i,j|g} -1 }}_{\Pr(\mtheta_g|\malpha_g)}  \prod_{g \in G} d\mtheta_g \\
&= \sum_{\omega \in \Omega} p_\omega \prod_{g \in G} \prod_{s_i \in S} \frac{1}{B(\valpha_{s_i|g})} \int \prod_{s_j \in S} \theta_{i,j|g}^{n_{i,j|g,\omega} + \alpha_{i,j|g} - 1}  d\mtheta_g \\
&= \sum_{\omega \in \Omega} \theta_\omega \prod_{g \in G} \underbrace{\prod_{s_i \in S} \frac{ B(\vn_{s_i|g,\omega} + \valpha_{s_i|g})}{B(\valpha_{s_i|g})}}_{\Pr(D_{g|\omega}|\malpha_g)} 
\end{align*}
This concludes the derivation of the marginal likelihood formula in \Cref{eq:ml-analytical}.

\clearpage
\section{Notation overview}
\label{app:notation}

The following table provides an overview of all important notations used throughout the article.

\bigskip
\begin{tabular}{l|p{10cm}}
    $S$ & set of all states $S = \{s_1, \ldots, s_n\}$\\
    $D$ & set of observed transitions $D = \{t_1, ..., t_m\}$ \\
    $G$ & set of all groups $G = \{g_1, \ldots, g_o\}$ \\
    $src_k,dst_k$ & the source state $src_k$ and the destination state $dst_k$ of transtion $t_k$ \\
    $i_k,j_k$ & the index of the source state $i_k$ and the destination state $j_k$ of transtion $t_k$ \\
    \hline
    
    $\gamma_{g|t}$ & probability for transition $t$ to belong to group $g$ \\
    
    $\vgamma_t$ & group assignment probabilities for a single transitions $\vgamma_t = \{ \gamma_{g|t} | g \in G \}$ \\
    $\mgamma$ & group assignment probabilities for all transitions $\mgamma = \{ \vgamma_t | t \in D \}$ \\
    \hline
    
    $\theta_{i,j|g}$ & probability of a transition from state $s_i$ to state $s_j$ for group $g$ \\
    $\vtheta_{s_i|g}$ & transition probabilities from state $s_i$ to all other states in group $g$, 
        \newline i.e., $\vtheta_{s_i|g} = (\theta_{i,1|g}, \ldots, \theta_{i,n|g})$ \\
    $\mtheta_g$ & transition probabilities between states for group $g$, 
        i.e., $\mtheta_g = \{ \vtheta_{s_i|g} ~|~ s_i \in S \}$ \\
    $\mtheta$ & transition probabilities for all groups $\mtheta = \{ \mtheta_g | g \in G \}$ \\
    \hline 
    $\mphi$ & belief in transition probabilities (from a hypothesis)\\
    $\phi_{i,j|g}$ & belief (from a hypothesis) in the probability of a transition from state $s_i$ to state $s_j$ for group $g$ \\
    $\alpha_{i,j|g}$ & Dirichlet parameter ($\in \mathbb{N}$) for the transition from state $s_i$ to state $s_j$ in group $g$ \\
    $\valpha_{s_i|g}$ & Dirichlet parameters for state $s_i$ in group $g$,
        i.e., $\valpha_{s_i|g} = (\alpha_{i,1|g}, \ldots, \alpha_{i,n|g})$ \\
    $\malpha_g$ & Dirichlet parameters for the transitions in group $g$, i.e., $\malpha_g = \{ \valpha_{s_i|g} ~|~ s_i \in S \}$ \\
    $\malpha$ & Dirichlet parameters for all groups $\malpha = \{ \malpha_g | g \in G \}$ \\
    \hline 
    
    $\Omega$ & the set of all group assignments $\Omega = \{\{(t_1, g_1), ..., (t_m, g_m)\} | (g_1, ..., g_m) \in G^{|D|}\}$ \\
    $\omega$ & a fixed group assignment $\omega \in \Omega$ for each transition in transition dataset $D$ \\
    $p_\omega$ & the probability for group assignment $\omega \in \Omega$ \\
    $n_{i,j|g,\omega}$ & the number of transitions in dataset $D$ from state $s_i$ to state $s_j$ given group $g \in G$ and group assignment $\omega \in \Omega$\\
    $\mn_{g,\omega}$ & the matrix $\mn_{g,\omega} = (n_{i,j|g,\omega})$ holds the number of transitions in dataset $D$ between all states given group $g \in G$ and group assignment $\omega \in \Omega$ 
\end{tabular}
% \end{table}

% \end{itemize}
\end{document}